\documentclass[aps,twocolumn,superscriptaddress]{revtex4-1}
\usepackage{graphics} % or graphicx
\usepackage{epsfig} % use this to include EPS figures & adjust figure size
\usepackage{amsmath}
\usepackage{color}

\begin{document}
%\title{Modulational instability analysis of Yukawa fluid under the Quasi-localization charged approximation (QLCA) framework}
\title{Modulational instability of a Yukawa fluid excitation under the Quasi-localization charged approximation (QLCA) framework}
\author{Sandip Dalui}
\email{dalui.sandip77@gmail.com}
\affiliation{Institute for Plasma Research, HBNI, Bhat, Gandhinagar 382428, India}
\author{Prince Kumar}
\author{Devendra Sharma}
\affiliation{Institute for Plasma Research, HBNI, Bhat, Gandhinagar 382428, India}
\affiliation{Homi Bhabha National Institute, Training School Complex, Anushaktinagar, Mumbai 400094, India}
\date{\today}

\begin{abstract}
%%%%%%%%%%%%%%%%%%%%%%%%%%%%%%%%%%%%%%%%%%%%%%%%
\noindent 
Collective response dynamics of a strongly coupled system departs from the
continuum phase upon transition to the quasicrystalline phase, or formation of
a Wigner lattice. The wave nonlinearity leading to the modulational
instability in recent studies, for example, of a quasicrystalline dusty 
plasma lattice, predicts inevitable emergence of macroscopic structures from 
mesoscopic carrier fluctuations. 
The modulational instability in the quasi crystalline or amorphous phase of
a strongly coupled system, uniquely accessed under the quasi-localized
charge approximation (QLCA), generates a narrower instability regime for 
entire spectral range. In comparison to the linear one dimensional chains of 
strongly coupled dust grains, the longitudinal modes for quasicrystalline 
phase show finite distinction in terms of the instability regime. The present 
QLCA based analysis shows system to be stable for arbitrarily long wavelength 
of perturbation for full range of screening parameter 
$\kappa=a/\lambda_{\rm D}$ beyond the value $\kappa=0.182$, where $a$ is the inter dust separation and $\lambda_{\rm D}$ is the plasma Debye length. 
%The present Qusai-localized charged approximation (QLCA) based model recovers a wide stable region except a small unstable region in a relatively small $\kappa$ value. 
However, this unstable region continuously grows with increase in the 
dust temperature which invoke the weak coupling effects.  
The present results show that as compared to the one dimensional chains, the 
more practical 2D and 3D strongly coupled systems are potentially stable 
with respect to the macroscopic amplitude modulations. The development of 
macroscopic structures from the mesoscopic fluctuations is therefore 
predicted to be rather restricted for strongly coupled systems with 
implications for systems where strongly coupled species are in a 
quasi-localized (semi-solid) phase.
\end{abstract}

%\pacs{}
%\pacs{52.25.Dg,52.35.Mw,52.35.Sb,52.65.Ff,94.05.Fg,94.05.Pt,94.20.wf,52.35.Fp}


\maketitle
%%%%%%%%%%%%%%%%%%%%%%%%%%%%%%%%%%%%%%%%%%%%%%%%%%%%%%%%%%%%%%%%%%%%%%%
% Beginning of the paper text
%%%%%%%%%%%%%%%%%%%%%%%%%%%%%%%%%%%%%%%%%%%%%%%%%%%%%%%%%%%%%%%%%%%%%%%
%%%%%%%%%%%%%%%%%%%%%%%%%%%%%%%%%%%%%%%%%%%%%%%%%%%%%%%%%%%%%%%%%%

\section{\label{Sec:1} Introduction}
The study of collective excitations and (or) various instabilities in  strongly coupled systems serve as a basis to get deep insight of many natural strongly coupled systems, such as white dwarf \cite{koester1986evolution}, neutron star \cite{kouveliotou2001neutron, chabrier2002dense}, as well as laboratory based systems, such as resonant side-bands \cite{shukla1996stimulated}, ultracold neutral plasma \cite{rosenberg2011instabilities}, classical 2D electron liquid trapped on the surface of liquid helium \cite{golden1992dielectric2d}, semiconductor electronic bilayers \cite{kalman1999collective}, polarized charged particles \cite{rosenberg1998effect} etc. Highly charged dust particles electrostatically suspended an electron-ion plasma, namely dusty plasma, serves as an accessibly laboratory system where dust potential energy exceeding the kinetic energy can be easily achieved \cite{merlino1998laboratory, fortov2005complex, shukla2015introduction}. Besides laboratories dusty plasmas appear in space \cite{horanyi1985trajectories, horanyi1986effects, goertz1989dusty, northrop1992dusty, tsytovich1997dust, shukla2015introduction} and various satellite observations \cite{whipple1981potentials, robinson1992spacecraft} confirm the existence of dusty (complex) plasmas in the neighbourhood of space stations and artificial satellites. 

A number of theoretical \cite{rosenberg1997dust, xie2000dust, anowar2009nonlinear, yaroshenko2010nonlinear, wang2016nonlinear} and experimental studies \cite{quinn2000experimental} of strongly coupled dusty plasma systems predict modifications in the linear and nonlinear collective excitation spectra. 
The modulational instability (MI) is widely treated in the context of dust acoustic 
waves in the existing literature \cite{amin1998modulational, kourakis2003modulational, kourakis2004oblique, duan2004modulational, misra2006modulational, el2006modulational, gill2010modulational, bains2013dust, khaled2021modulational}.
Among the studies addressing the modulational instability in strongly coupled systems, a limited number has admitted the explicit localization by treating a linear one dimensional chain of dust grains  \cite{amin1998amplitude, amin1998modulational_dustlattice, kourakis2006nonlinear}
 whereas in other studies a more indirect inclusion of strong coupling is done by letting an effective dust temperature \cite{sultana2020dust} represent the strong coupling effect \cite{ikezi1986coulomb, thomas1994plasma, chu1994direct, misawa2001experimental, xie2002modulational, chaudhuri2019solitary, el2015modeling, chaudhuri2019solitary}. 
Both these approaches treat intrinsically one-dimensional setups.
However, the dust localization is represented by a more general spherically symmetric pair correlation function in a quasi-localized charge approximation (QLCA) approach of strongly coupled systems. The QLCA effectively accounts for the regime about the melting point and transient amourphous phases.
Various instabilities such as ion-dust instability \cite{rosenberg1998effect, kalman2003instabilities}, resonant \& Buneman-type instability \cite{kalman2003instabilities}, dust-dust instabilities \cite{rosenberg2012instabilities} and dust acoustic (DA) instability \cite{rosenberg2012instabilities, rosenberg2014effect} have also been investigated under the QLCA formalism. 
The MI is however not treated under the QLCA approach in the exciting 
literature to best of our knowledge. 

Relevant to many modern laboratory applications discussed below, the present 
study shows that in comparison to the results from MI in 
one-dimensional dust lattice excitations, the regime of instability can be 
highly restricted in a general spherically symmetric dust structure described 
by a more general pair-correlation function $g(r)$. In particular, it is 
shown that for wide range of values of screening parameter ($\kappa=a/\lambda_{D}$) explored, 
the unstable region is restricted upto a rather small value of $\kappa$. 
For strong coupling limit, the parameter space explored by variation of both 
screening parameter ($\kappa$) and the dust temperature ($\sigma_{d} = T_{d}/Z_{d} T_{i}$) it is shown 
that the temperature enhances the dimension of unstable zone in the parameter 
space. The stabilization of amplitude modulation is verified for larger 
values of $k$ by finding the maximum growth rate of the instability and showing that it indeed reduces to zero at the small value of $\kappa=0.183$ 
%for $\sigma_{d}=0.000353 Z_{d}T_{i}$, where $Z_{d}$ is the dust charge and 
for the dust temperature $T_{d}=3.53\times 10^{-4} Z_{d}T_{i}$, 
where $Z_{d}$ is the dust charge and $T_{i}$ is the ion temperature. 
%in the range from 0.13 to 0.16 and as we already predicts that the MMGRI become zero at $\kappa$ = 0.183.

We additionally have dust temperature as a parameter which extends the 
relevance of our analysis to systems where instability thresholds can be 
strong function of the temperature of the trapped species. Examples include 
the edge of the stability region of an RF/Laser ion trap 
\cite{zhou2021mass} where the instabilities 
arising from collective excitation of lattice ions can facilitate ion 
manipulation as the ratio between the collective interaction energy of ions 
($\sim 0.1$ eV) and depth of RF trapping field ($> 100$ eV) drops 
significantly from its extreme bulk value of $> 10^{-4}$. The similar 
collective interactions are involved in entrainment arising from Bragg 
scattering of unbound neutrons by the collective excitement of the Coulomb 
lattice in the inner crust of a neutron star \cite{chamel2016collective}. 
A number of other examples
with either positive or negative consequences of the instability can also be 
cited from fusion plasmas \cite{stacey2007survey}, ultracold neutral 
plasmas \cite{killian2007ultracold, lyon2016ultracold}. The effect of both 
coupling and the temperature remains important on the instability threshold 
and are incorporated in the present treatment showing that the instability 
threshold does reappear with the change of dust temperature while it is found to be
stable over the entire scale spectrum at lower temperatures in contrast to 
the results for linear one-dimensional chain \cite{amin1998amplitude, amin1998modulational_dustlattice, kourakis2006nonlinear}.

%The rouge waves (Tsunami) may generates due to the effect of modulation instability in strongly coupled system \cite{guo2016modulation}. 
%Modulation instability may be found in the optical fiber \cite{tai1986observation, shukla1986modulational}. 
%Modulation is one of the powerful phenomenon to control the fusion thermal instability in Tokamak Plasmas \cite{stacey2007survey}. 
%Modulation plays a significant role to control the fire in Fusion Reactor \cite{hui1994robust} and Compact Ignition Tokamak (CIT) \cite{chaniotakis1989cit}.

This paper is organized as follows. 
In Sec.~\ref{fluid_eqn}, the QLCA based analytical fluid model is considered, and consequently, the linear dispersion relation is derived. 
In Sec.~\ref{Sec_NLSE}, using the reductive perturbation method (RPM) \cite{taniuti1969perturbation, asano1969perturbation}, the spatiotemporal nonlinear 
Schr{\"o}dinger equation (NLSE) is derived within the QLCA framework.
In Sec.~\ref{sec:Modulation_instability}, the nonlinear dispersion relation of modulated wave and the maximum modulational growth rate of instability (MMGRI) are analytically derived. 
Results and discussions on instability analysis of modulated wave are presented in Sec.~\ref{Results}. 
The summary and conclusions are presented in Sec.~\ref{Conclusions}.

\section{\label{fluid_eqn} Derivation of spatiotemporal equations with in the QLCA framework}
%%%%%%%%%%%%%%%%%%%%%%%%%%%%%%%%%%%%%%%%%%%%%%%%%%%%%%%%%%%%%%%%%%%%%%
%\section{Basic Equations}
We start with a more general expression described in recent analysis on  strongly coupled rotating dusty plasma under the QLCA framework \cite{kumar2021collective} 
%The QLCA theory has been successfully applied to study collective excitation of the liquid phase strongly coupled systems in a rotating \citep{kumar2021collective} as well as non-rotating frame\citep{Golden}. The microscopic equation of motion of the dust particle in a rotating frame,  for the component $r_{i\mu}$ aligned to the direction $\mu~(=x,y)$,
%% 
 \begin{eqnarray}
\begin{split}
&& m_{\rm d}\frac{\partial^{2}{r}_{i\mu}}{\partial t^2}=
	\sum_{j} {K}_{ij\mu\nu}r_{j\nu}
	-2m_{\rm d}\left[{\bf \Omega_{r}} \times \frac{\partial {\bf r}_{i}}{\partial t}\right]_{\mu}\\
&& -m_{\rm d}[{\bf \Omega_{r}}\times({\bf \Omega_{r}}\times {\bf r}_{i})]_{\mu} 
	-\frac{\partial V}{\partial r_{\mu}}=0,
\end{split}
	\label{particle-eq_rotating}
\end{eqnarray} 
where the second and third terms in the right-hand side are the Coriolis force and centrifugal force, respectively.
%
%The quantity V is the dust confinement potential whose  gradient balances the corresponding component of the centrifugal force in the typical equilibrium condition \citep{kumar2021collective}.
This equation can be reduced, in a non-rotating frame ($\Omega_{r} \rightarrow 0$), to a form given as \citep{golden2000quasilocalized} 
 \begin{eqnarray}
\begin{split}
m_{\rm d}\frac{\partial^{2}{r}_{i\mu}}{\partial t^2}=
	\sum_{j} {K}_{ij\mu\nu}r_{j\nu},
\end{split}
	\label{particle-eq}
\end{eqnarray}
where the non-retarded limit of ${K}_{ij\mu\nu}$ defines the potential 
energy of the strongly coupled Yukawa fluid. The standard QLCA prescription \cite{golden2000quasilocalized} led to the linearized version of the above  equation in their spectral space, given as,
\begin{eqnarray}
\begin{split}
	-m_{\rm d}\omega^2
	\frac{1}{\sqrt{Nm_{\rm d}}}\xi_{{\bf k}\mu}(\omega)\\
	-\frac{N}{V_{dD}}
	k_{\mu}k_{\nu}\psi_{dD}({\bf k},\omega)
	\frac{1}{\sqrt{Nm_{\rm d}}}\xi_{\bf k\nu}(\omega)\\
	-\frac{1}{V_{dD}}\sum_{\bf q} 
	q_{\mu}q_{\nu}\psi_{dD}({\bf q},\omega)\\
	\times [S(|{\bf k-q}|)-S(|{\bf q}|)]
	\frac{1}{\sqrt{Nm_{\rm d}}}\xi_{\bf k\nu}(\omega)\\
%
%	-\frac{1}{V_{dD}}\sum_{\bf q,p} 
%	q_{\mu}q_{\nu}\psi_{dD}({\bf q},\omega)
%	[-N^{2}\delta_{\bf q}\delta_{\bf k-p}]
%	\frac{1}{\sqrt{Nm_{d}}}\xi_{\bf p\nu}(\omega)\\
%
%	-\frac{1}{V_{dD}}\sum_{\bf q} 
%	q_{\mu}q_{\nu}\psi_{dD}({\bf q},\omega)
%	[-NS(|{\bf q}|)]
%	\frac{1}{\sqrt{Nm_{\rm d}}}\xi_{\bf k\nu}(\omega)\\
%
%	-\frac{1}{V_{dD}}\sum_{\bf q,p} 
%	q_{\mu}q_{\nu}\psi_{dD}({\bf q},\omega)
%	[+N^{2}\delta_{\bf q}\delta_{\bf k-p}]
%	\frac{1}{\sqrt{Nm_{\rm d}}}\xi_{\bf p\nu}(\omega)\\
%
%	+2i\omega m_{\rm d}~\epsilon_{\nu\delta\mu} \Omega_{\nu}
%	\frac{1}{\sqrt{Nm_{\rm d}}}\xi_{\bf k\delta}(\omega)\\
%	=Ze\sum_{\bf q}E_{\bf q}(\xi_{\bf q\mu},\omega)n_{k-q}
	%=ZeE_{\bf k}(\xi_{\bf k\mu},\omega).
=0.
\end{split}
\label{fourier_space_equation}
\end{eqnarray}   

In order to analyze the nonlinear effect within the QLCA framework, we require that the averages are done over spatiotemporal functions rather than their Fourier transformations\cite{Prince 3red}. In the most approximate approach, we let the fluid conservation equations represent the evolution of these ensemble averages. 
%
%The central idea of localization involves using the macroscopic variables 
%obtained from the ensemble averaged particle equations. For a nonlinear 
%approach it is however required that the averages are done over spatiotemporal
%functions rather than their Fourier transformations. 
%We however acknowledge the presence of
%a spatial ordering of the dust sites by allowing the dynamical matrix 
%$D_{L,T}$ (determining mechanical response of the system) to be computed 
%via the grain-grain correlation energy which changes based on strain in 
%spatial ordering in addition to the response because of associated 
%background plasma compression.
%
This ensemble averaged (macroscopic) momentum equation is  
%Eq.~(\ref{momentum-balance-rpa})
%and, when the dust kinetic energy is once again neglected because of being 
%a few orders smaller than the representative strong coupling term  
%$\frac{\partial P_{di}}{\partial x}$, can be written as,
% 
%We consider the continuity equation, the equation of motion \cite{hou2004theoretical,hou2009wave} and the equation of state for the dust fluid are 
%--------------------------------------------------------------------------------------------------------------------------------------------------------
%
%--------------------------------------------------------------------------------------------------------------------------------------------------------
\begin{eqnarray}\label{Equation_of_motion}
 \frac{\partial u_{d}}{\partial t}+u_{d}\frac{\partial u_{d}}{\partial x} &=& \frac{e Z_{d}}{m_{d}}\frac{\partial \phi_{d}}{\partial x} - \frac{1}{m_{d}n_{d}} \Bigg( \frac{\partial P_{di}}{\partial x}\Bigg) \nonumber \\
  && - \frac{1}{m_{d}n_{d}} \Bigg( \frac{\partial P_{dk}}{\partial x} \Bigg)  .
\end{eqnarray}

Similarly, the continuity equation of macroscopic particles obtained by ensemble averaging over the dust sites is 

\begin{equation}\label{Equation_of_continuity}
\frac{\partial n_{d}}{\partial t}+\frac{\partial }{\partial x}(n_{d}u_{d}) = 0  ,
\end{equation}
 and the equation of state is
\begin{eqnarray}\label{Equation_of_pressure}
 P_{dk} = P_{k0} n_{d}^{\gamma} .
\end{eqnarray}
%
%To study the modulational instability of dust acoustic (DA) waves in strongly coupled dusty plasmas, we consider the complex plasma composed of Maxwell-Boltzmann distributed ions and electrons, and negatively charged dust grains.
Where, $n_{d}$, $u_{d}$, $\phi_{d}$ and $P_{dk}$ are the number density, velocity, the electrostatic potential and the pressure of dust particles, respectively. And also, $m_{d}$ and $Z_{d}$ are the mass of a dust particle and the average number of electrons residing on a dust particle, respectively. 
%%%%%%%%%%%%%%%%%%%%%%%%%%%%%%%%%%%%%%%%%%%%%%%%%%%%%%%%%%%%%%%%%%%%%%%%%%%%%%%%%%%%
% The inter-particle interactions of dust particles within the short-range result the strongly coupling in the system, therefore,
Local field effects are introduced via a correction $P_{di}$ to the ideal-gas pressure term \cite{hou2004theoretical, hou2009wave} and this correction 
$P_{di}$ contains essential structural information.
So, according to Hou \textit{et al.}, \cite{hou2004theoretical,hou2009wave}, we consider 
\begin{eqnarray}\label{pdi}
 \frac{\partial P_{di}}{\partial x} = \Bigg( \frac{\delta P_{di}}{\delta n_{d}} \Bigg)_{T_{d}}  \frac{\partial n_{d}}{\partial x} ,
\end{eqnarray}
and the dust layer compressibility, $ \alpha \equiv \big( \frac{\delta P_{di}}{\delta n_{d}} \big)_{T_{d}} $, is directly related to the dust-dust correlation energy of the Yukawa system \cite{lado1978hypernetted,hartmann2005equilibrium}. 
% 
%This dust layer compressibility $ \alpha$ is directly related to the system’s correlation energy can be considered \cite{hou2004theoretical,hou2009wave} as 
%\begin{eqnarray}\label{alpha_2}
  %\alpha = \frac{n_{d0}}{m_{d}} \frac{\partial^2 \Big( n_{d0} \upsilon_{c}(n_{d0}) \Big)}{\partial n_{d0}^2}   ,
%\end{eqnarray}
%where $\upsilon_{c}(n_{d0})$ be the correlation energy \cite{hou2004theoretical,hou2009wave} of the Yukawa system. 
%Although, Hou \textit{et al.} \cite{hou2004theoretical} have considered the expression of $\upsilon_{c}$ explicitly as numerically derived by Lado \cite{lado1978hypernetted}. But, here we will consider this correlation energy according to the QLCA approach as considered by Khrapak \textit{et al.} \cite{khrapak2016long}.

%-------------------------------------------------------------------------------------------------------------------------------------------------------
 The system of basic equations (\ref{Equation_of_motion}) - (\ref{Equation_of_pressure}) are closed by the following Poisson equation, 
\begin{eqnarray}\label{Equation_of_Poisson}
\frac{\partial^{2} \phi_{d}}{\partial x^{2}} = 4\pi e (n_{e} - n_{i} + Z_{d}n_{d}) ,
\end{eqnarray}
where $n_{i}$ and $n_{e}$ are the ion and electron number density, respectively, follow the Boltzmann distribution give as,
\begin{eqnarray}\label{ne_and_ni}
  n_{e} = n_{e0} \displaystyle \exp \Big[{\frac{e\phi_{d}}{K_{B}T_{e}}}\Big],  ~ n_{i} = n_{i0} \displaystyle \exp \Big[-{\frac{e\phi_{d}}{K_{B}T_{i}}}\Big] ,
\end{eqnarray}
 $T_{i}$ and $T_{e}$ are the ion and electron temperature, respectively.
The charge neutrality condition is given as
\begin{eqnarray}\label{neutrality_condition}
  n_{e0} - n_{i0} + Z_{d}n_{d0} =0 .
\end{eqnarray}

\subsection{Linear Theory}
 We solve a full set of nonlinear equations (\ref{Equation_of_motion}-\ref{neutrality_condition}) by  perturbing the physical variables with small amplitude of the form $e^{i(kx - \omega t)}$, in order to recover the linear dispersion relation, which is given below, of the dust acoustic wave in the strongly coupled dusty plasma, 
\begin{eqnarray}\label{LDR1}
  \omega^2 (k)= \omega_{0}^{2} (k) + \frac{\gamma v_{th}^2}{2} k^2 + \alpha k^2 ,
\end{eqnarray}
where
\begin{eqnarray}\label{w0}
 \omega_{0}^{2} (k) = \frac{\omega_{pd}^2 k^2}{k^2 + \frac{1}{\lambda_{D}^2}} , ~~  \omega_{pd}^2 = \frac{4\pi e^2 n_{d0} Z_{d}^2}{m_{d}}  ,
\end{eqnarray}
%\begin{eqnarray}\label{wpd}
%\omega_{pd}^2 = \frac{4\pi e^2 n_{d0} Z_{d}^2}{m_{d}}  ,
%\end{eqnarray}
\begin{eqnarray}\label{lambda_D}
  \frac{1}{\lambda_{D}^2} = \frac{1}{\lambda_{De}^2} + \frac{1}{\lambda_{Di}^2} = \frac{4\pi e^2 n_{e0}}{K_{B}T_{e}} + \frac{4\pi e^2 n_{i0}}{K_{B}T_{i}} , 
\end{eqnarray}
\begin{eqnarray}\label{v_th}
  v_{th}^2 = \frac{2K_{B}T_{d}}{m_{d}} .
\end{eqnarray}
%The results obtained from the semi-analytical approach essentially coincide \cite{hou2004theoretical} with the results of quasi-localized charged approximation (QLCA) approach for the long-wavelength, as $ \alpha  = \lim_{k \to 0} \frac{D_{L}(k)}{k^2} $ where the QLCA dynamical matrix $D_{L}(k)$ is given by \cite{khrapak2016long}

The last term of the equation (\ref{LDR1}), i.e., $\alpha k^2$  incorporates the essential the QLCA  (strong coupling) effects in the formulation since the  dust layer compressibility ($\alpha$) can be approximated by the QLCA dynamical matrix $D_{L}$ \cite{hou2009wave, khrapak2016long}, in a long wavelength limit, as $\displaystyle \alpha  = \lim_{k \to 0} \frac{D_{L}(k)}{k^2} $ \cite{hou2009wave} where, 
%
%an initial form of a local field correction (LFC) . 
%
%The dust layer compressibility $\alpha $ can be considered from the molecular dynamical simulation \cite{hartmann2005equilibrium}, 
%
%
%Hou \textit{et al.} \cite{hou2004theoretical} have considered the expression of $\upsilon_{c}$ explicitly as numerically derived by Lado \cite{lado1978hypernetted}. But, here we will consider this correlation energy according to the QLCA approach as considered by Khrapak \textit{et al.} \cite{khrapak2016long}.
%
%The dust layer compressibility $\alpha $ is approximated by the QLCA dynamical matrix $D_{L}(k)$ \cite{hou2009wave, khrapak2016long} for a long wavelength limit given as $\displaystyle \alpha  = \lim_{k \to 0} \frac{D_{L}(k)}{k^2} $ where 
\begin{eqnarray}\label{DL} 
  D_{L} (k) = && - \omega_{0}^{2} (k) + e^{-R\kappa}  \Bigg\{ \big( 1+ R\kappa \big) \nonumber \\
	&& \times \Bigg( \frac{1}{3}  - \frac{2\cos {(Rk)} }{(Rk)^2} + \frac{2\sin {(Rk)}}{(Rk)^3} \Bigg) \nonumber \\
  && - \frac{\kappa^2}{k^2 + \kappa^2}  \Bigg( \cos {(Rk)}+ \frac{\kappa \sin {(Rk)}}{k} \Bigg)\Bigg\}  , 
\end{eqnarray}
 $R$ $\approx$ 1+ $\kappa$/10  is an excluded volume \citep{khrapak2016long} and $\kappa = \frac{a}{\lambda_{D}} $ is the screening parameter. After substituting the parameter $\alpha $ in eq.(\ref{LDR1}), the equation becomes, 
%
% For long-wavelength limit, substituting the dust layer compressibility $\alpha $ into the linear DR (\ref{LDR1}), we get 
\begin{eqnarray}\label{LDR31}
   \omega^2 (k) = \omega_{0}^2 (k) + \frac{\gamma v_{th}^2}{2} k^2 +  D_{L}(k)  .
\end{eqnarray}
 The eq.(\ref{LDR1}) represents the linear dispersion relation of dust acoustic wave in the strongly coupled Yukawa system.
\subsection{Normalized Basic Equations:}

 After normalizing the system of equations (\ref{Equation_of_motion}) - (\ref{Equation_of_pressure}) and (\ref{Equation_of_Poisson}), we get the following normalized equations:
\begin{equation}\label{Equation_of_continuity_1}
\frac{\partial \bar{n}_{d}}{\partial t}+\frac{\partial }{\partial x}(\bar{n}_{d}\bar{u}_{d}) = 0 ,
\end{equation}
%------------------------------------------------------------------
\begin{equation}\label{Equation_of_motion_1}
\frac{\partial \bar{u}_{d}}{\partial t}+ \bar{u}_{d}\frac{\partial \bar{u}_{d}}{\partial x} = \mu \frac{\partial \bar{\phi}_{d}}{\partial x} - \mu \gamma\sigma_{d} \bar{n}_{d}^{\gamma-2} \frac{\partial \bar{n}_{d}}{\partial x} - \frac{{\alpha}}{\bar{n}_{d}} \frac{\partial \bar{n}_{d}}{\partial x},  
\end{equation}
%------------------------------------------------------------------
\begin{equation}\label{Equation_of_Poisson_1}
\frac{\partial^{2} \bar{\phi_{d}}}{\partial x^{2}} = \frac{1}{\mu} [ (\bar{n}_{d} - 1) + h_{1} \bar{\phi}_{d} + h_{2} \bar{\phi}_{d}^2 + h_{3} \bar{\phi}_{d}^3 ] .
\end{equation}
The space variable ($x$) and the time ($t$) are respectively normalized by $a $ and $\omega_{pd}^{-1} = \sqrt{\frac{m_{d}}{4\pi e^2 n_{d0}Z_{d}^2}}$.  The dust number density ($\bar{n}_{d}$), the dust velocity  ($\bar{u}_{d}$) and dust electrostatic potential ($\bar{\phi}_{d}$) are normalized by $n_{d0}$, $a / \omega_{pd}^{-1}$ and $ \frac{K_{B} T_{i}}{e},  $ respectively. Other parameters involved in the calculation are given as, $\displaystyle \sigma_{d}=\frac{T_{d}}{Z_{d}T_{i}} $, $\displaystyle \mu = \frac{c_{s}^2}{a^2 \omega_{pd}^2} $ and $\displaystyle c_{s} = \sqrt{\frac{Z_{d}K_{B} T_{i}}{m_{d}}}$.
% and $\frac{1}{\lambda_{D}^2} = \frac{1}{\lambda_{De}^2} + \frac{1}{\lambda_{Di}^2} = \frac{4\pi e^2 n_{e0}}{K_{B}T_{e}} + \frac{4\pi e^2 n_{i0}}{K_{B}T_{i}} $.
 %
 The charge neutrality condition  (\ref{neutrality_condition}) can be written as
\begin{eqnarray}\label{neutrality_condition_1}
  \overline{N}_{e0} - \overline{N}_{i0} + 1 = 0, 
\end{eqnarray}
where $\overline{N}_{e0} = \frac{n_{e0}}{Z_{d}n_{d0}}$ and $\overline{N}_{i0} = \frac{n_{i0}}{Z_{d}n_{d0}}$. 

The number densities (\ref{ne_and_ni}) of both the electrons ($\bar{n}_{e}$) and ions ($\bar{n}_{i}$) can be written as
\begin{eqnarray}\label{ne_and_ni_1}
  \bar{n}_{e} = \overline{N}_{e0} \displaystyle \exp \Big[\sigma_{ie}\bar{\phi}_{d}\Big],  ~ \bar{n}_{i} = \overline{N}_{i0} \displaystyle \exp \Big[-\bar{\phi}_{d}\Big] , 
\end{eqnarray}
where 
\begin{eqnarray}
	\sigma_{ie} =\frac{T_{i}}{T_{e}} ,  
\end{eqnarray}
\begin{eqnarray}
  h_{1} = \overline{N}_{e0} \sigma_{ie} + \overline{N}_{i0} ,
\end{eqnarray}
\begin{eqnarray}
  h_{2} = \frac{1}{2} \bigg( \overline{N}_{e0} \sigma_{ie}^2 - \overline{N}_{i0} \bigg) ,
\end{eqnarray}
\begin{eqnarray}
	h_{3} = \frac{1}{6} \bigg( \overline{N}_{e0} \sigma_{ie}^3 + \overline{N}_{i0} \bigg) .
\end{eqnarray}

%%%%%%%%%%%%%%%%%%%%%%%%%%%%%%%%%%%%%%%%%%%%%%%%%%%%%%%%%%%%%%%%%%%%%%
%\section{Derivation of the NLSE}
%%%%%%%%%%%%%%%%%%%%%%%%%%%%%%%%%%%%%%%%%%%%%%%%%%%%%%%%%%%%%%%%%%%%%%
\section{\label{Sec_NLSE}  Derivation of spatiotemporal Nonlinear Equation}

A nonlinear Schr{\"o}dinger equation is derived to study the MI of DA waves in strongly coupled Yukawa system within the QLCA framework. Now, to derive the NLSE, we consider the following stretchings  
%-----------------------------------------------------------------------------------------------
\begin{eqnarray}\label{stretchings}
 \xi = \epsilon (x-V_{g}t) ,~ \tau=\epsilon^{2}t ,
\end{eqnarray}
%----------------------------Perturbation ----  Perturbation ----
and the perturbation expansions of dependent variables are
\begin{eqnarray}\label{Perturbation_of_f}
  \bar{f}_{d} = f^{(0)} + \sum_{l=1}^{\infty} \epsilon^{l} ~ \sum_{a=-\infty}^{+\infty} f_{a}^{(l)}(\xi,\tau) ~ e^{ia (kx-\omega t)}  ,
\end{eqnarray}
where $k$ is carrier wave number, $\omega$ is carrier  wave frequency and $\bar{f}_{d}$ = $\bar{n}_{d}$, $\bar{u}_{d}$, $\bar{\phi}_{d}$ with $f^{(0)} = [1~ 0~ 0]^{T}$. 
% $n^{(0)}=1$, $u^{(0)}=\phi^{(0)}=0$. 

Putting the perturbation expansions (\ref{Perturbation_of_f}) for the field quantities $\bar{n}_{d}$, $\bar{u}_{d}$, $\bar{\phi}_{d}$, into the set of Yukawa fluid equations (\ref{Equation_of_continuity_1}) - (\ref{Equation_of_Poisson_1}), and sorting the distinct equations of distinct powers of $\epsilon$, we get a sequence for distinct orders $l=1,2,3, \cdots$ and a sub-sequence for distinct harmonics $a=0, \pm 1, \pm 2, \cdots$.

\subsection{First order $O(\epsilon=1)$:}
The set of zeroth harmonic ($a=0$) equations for the system of basic equations (\ref{Equation_of_continuity_1}) - (\ref{Equation_of_Poisson_1}) are identically satisfied \cite{dalui2017modulational}.  

Solving the first harmonic ($a=1$) equations of the system of basic equations (\ref{Equation_of_continuity_1}) - (\ref{Equation_of_Poisson_1}), we obtained the linear DR of DA waves for strongly coupled Yukawa system 
\begin{eqnarray}\label{dispersion_relation_1}
 \omega^{2} = \frac{k^2}{k^{2}+ \kappa^2} + \gamma \sigma_{d} k^2 \mu + \alpha k^2 ,
\end{eqnarray}
%where the screening parameter is $\displaystyle \kappa=\frac{a} {\lambda_{D}} \Longleftrightarrow \kappa^2 = \frac{h_{1}}{\mu} $.
where $\displaystyle \kappa^2 = \frac{h_{1}}{\mu} $.

%The semi analytical approximation essentially coincides with the quasi-localized charged approximation (QLCA) for the long-wavelength \cite{kalman2004two, hou2009wave}, then $\alpha  = \lim_{k \to 0} \frac{D_{L}(k)}{k^2} $, where the QLCA dynamical matrix $D_{L}(k)$ \cite{golden2000quasilocalized, kalman2000collective, khrapak2016long} is given by 
%Now, the isothermal dust layer compressibility $\alpha$ \cite{hou2009wave} is approximated by the QLCA dynamical matrix $D_{L}(k)$ \cite{khrapak2016long} for the long-wavelength as $\displaystyle \alpha  = \lim_{k \to 0} \frac{D_{L}(k)}{k^2} $ where 
 
%\begin{eqnarray}\label{DL1}
  %D_{L} (k) = && \Bigg[ - \frac{k^2}{k^2 + \kappa^2} + e^{-R\kappa}  \Bigg\{ \big( 1+ R\kappa \big) \nonumber \\
	%&& \times \Bigg( \frac{1}{3}  - \frac{2\cos {(Rk)} }{(Rk)^2} + \frac{2\sin {(Rk)}}{(Rk)^3} \Bigg) \nonumber \\
  %&& - \frac{\kappa^2}{k^2 + \kappa^2}  \Bigg( \cos {(Rk)}+ \frac{\kappa \sin {(Rk)}}{k} \Bigg)\Bigg\} \Bigg]  .  
%\end{eqnarray}
%
%Now, substituting the expression of $D_{L} (k)$ as given in equation (\ref{DL1}), in to the linear dispersion relation (\ref{dispersion_relation_1}), we get the linear dispersion relation for strongly coupled dusty plasmas: 
%\begin{eqnarray}\label{LDR21}
  %\omega^2 &=& \gamma \sigma_{d} k^2  \mu + \Bigg[ e^{-R\kappa}  \Bigg\{ \big( 1+ R\kappa \big) \nonumber \\
	%&& \times \Bigg( \frac{1}{3}  - \frac{2\cos {(Rk)} }{(Rk)^2} + \frac{2\sin {(Rk)}}{(Rk)^3} \Bigg) \nonumber \\
  %&& - \frac{\kappa^2}{k^2 + \kappa^2}  \Bigg( \cos {(Rk)}+ \frac{\kappa \sin {(Rk)}}{k} \Bigg)\Bigg\} \Bigg] .
%\end{eqnarray}

If we consider the case for a weakly coupled limit of dusty plasma, i.e., $R \rightarrow 0$ then the QLCA dynamical matrix $D_{L}(k) \to 0$ as well as the isothermal dust layer compressibility $\alpha \to 0$, and consequently, the linear DR (\ref{dispersion_relation_1}) of a strongly coupled dusty plasma reduces to the conventional DA wave linear DR \cite{khrapak2017practical} as follows 
\begin{eqnarray}\label{conventional_DA_LDR}
 \omega^{2} = \frac{k^{2}}{k^{2} + \kappa^{2}} + \gamma \sigma_{d} \mu k^{2} . 
\end{eqnarray}

\subsection{Second order $O(\epsilon=2)$:}
\subsubsection{First Harmonics ($a=1$)}
Solving the set of first harmonic equations of the basic equation, we get the following compatibility condition which refers to the group velocity 
\begin{eqnarray}\label{eqn for Vg}
  V_{g} = \frac{\omega^{2}- W_{d}^{4}}{\omega k}= \frac{k}{\omega}  \Bigg[\frac{h_{1} \mu}{(k^{2} \mu+h_{1})^{2}} + \mu\gamma \sigma_{d} +  \alpha \Bigg],
\end{eqnarray}
where $W_{d}^{2}=\omega^{2}- \Big(\mu\gamma\sigma_{d} +  \alpha \Big) k^{2}$.
%where we have used the linear DR (\ref{dispersion_relation_1}) to simplify the equation (\ref{eqn for Vg}).

\subsubsection{Second Harmonic ($ a=2 $)}
Solving the set of second harmonic equations of the basic equations, we get 
\begin{eqnarray}\label{form_of_phi22}
  \Big(\phi_{2}^{(2)},n_{2}^{(2)},u_{2}^{(2)} \Big) = \Big(A_{\phi_{d}} , A_{n_{d}} , A_{u_{d}} \Big) [\phi_{1}^{(1)}]^{2} ,
\end{eqnarray}
where
\begin{eqnarray}
 A_{\phi_{d}} = - \Bigg[ \frac{h_{2}}{3k^{2} \mu} + \frac{k^{2}\omega^{2} \mu}{2W_{d}^{6}} + (\mu\gamma \sigma_{d} g_{1} - \alpha ) \frac{k^{4} \mu}{6W_{d}^{6}} \Bigg] ,\label{form_of_A_phi} 
 \end{eqnarray}
\begin{eqnarray}
 A_{n_{d}} = -\Big[ (4k^{2} \mu + h_{1}) A_{\phi_{d}} + h_{2} \Big] ,\label{form_of_A_n}
 \end{eqnarray}
\begin{eqnarray}
 A_{u_{d}} = \frac{\omega}{k} \Bigg[ A_{n_{d}} - \frac{k^4}{W_{d}^4} \mu^2 \Bigg],\label{form_of_A_w}
\end{eqnarray}
and $g_{1}=(\gamma - 2) $.

\subsubsection{Zeroth Harmonic ($ a=0 $)}
Solving the set of zeroth harmonic equations of the basic equations, we get 
\begin{eqnarray}\label{form_of_phi20}
  \Big(\phi_{0}^{(2)},n_{0}^{(2)},u_{0}^{(2)}\Big) = \Big(B_{\phi_{d}} , B_{n_{d}} , B_{u_{d}}\Big) |\phi_{1}^{(1)}|^{2} ,
\end{eqnarray}
where
\begin{eqnarray}\label{form_of_B_phi}
   B_{\phi_{d}} &=& - \frac{ 1 }{W_{d}^{4}[h_{1} \{V_{g}^{2}-(\mu\gamma \sigma_{d} + \alpha)\}- \mu]} \times \nonumber \\
	&& \Bigg[ \mu^2 \Big\{( \mu\gamma \sigma_{d}g_{1} - \alpha) k^{4} + k^{2} \omega (2 k V_{g}+\omega)\Big\} \nonumber \\ 
	&& + 2 h_{2} W_{d}^{4} \{V_{g}^{2}-( \mu\gamma \sigma_{d} + \alpha)\} \Bigg] ,
\end{eqnarray}
\begin{eqnarray}\label{form_of_B_n}
  B_{n_{d}} =  - \bigg[h_{1} B_{\phi_{d}} + 2h_{2} \bigg] ,
\end{eqnarray}
\begin{eqnarray}\label{form_of_B_w}
  B_{u_{d}} = V_{g} B_{n_{d}} - \frac{2k^{3}\omega }{W_{d}^{4}} \mu^2  .
\end{eqnarray}

\subsection{Third Order ($ l=3 $) : First Harmonic ($ a=1 $)}
Solving the set of first harmonic equations of the basic equations, we get the following NLSE 
\begin{eqnarray}\label{NLSE}
 i \frac{\partial \phi_{1}^{(1)}}{\partial \tau} + P_{d} \frac{\partial^{2} \phi_{1}^{(1)}}{\partial \xi^{2}} + Q_{d} |\phi_{1}^{(1)}|^{2} \phi_{1}^{(1)} = 0,
\end{eqnarray}
where
\begin{eqnarray}\label{expression_of_P}
 P_{d} &=& - \frac{W_{d}^{4}}{2k^{2}\omega} \Bigg[ 1- \frac{k^4}{W_{d}^6} \Big(V_{g}-\frac{\omega}{k}\Big)\Big(3V_{g}\frac{\omega^2}{k^2} \nonumber \\
&& - 3( \mu\gamma\sigma_{d}+ \alpha)\frac{\omega}{k} - \frac{\omega^3}{k^3} + ( \mu\gamma\sigma_{d}+\alpha) V_{g}\Big) \Bigg],
\end{eqnarray}
\begin{eqnarray}\label{expression_of_Q}
 && Q_{d} = - \frac{W_{d}^{4}}{2k^{2}\omega \mu} \Bigg[ 2 \frac{k^{3}\omega \mu}{W_{d}^4}(A_{u_{d}}+B_{u_{d}}) + \nonumber \\ 
 && \frac{k^{2}\mu}{W_{d}^{4}}\{\omega^{2} + (\mu\gamma\sigma_{d}g_{1}- \alpha)\}k^{2})(A_{n_{d}}+B_{n_{d}})\nonumber \\ 
 && + (\mu\gamma\sigma_{d}g_{2} + \alpha) \frac{k^{8} \mu^3}{W_{d}^8} - 3h_{3} -2h_{2}(A_{\phi_{d}}+B_{\phi_{d}}) \Bigg].
\end{eqnarray}
The $P_{d}$ and $Q_{d}$ are the coefficients of the dispersive and nonlinear term in the NLSE, respectively. The QLCA effects via $\alpha$ is appearing in the expression of $P_{d}$ and $Q_{d}$ which are being explored in the next section.
%%%%%%%%%%%%%%%%%%%%%%%%%%%%%%%%%%%%%%%%%%%%%%%%%%%%%%%%%%%%%%%%%%%%%%
\section{\label{sec:Modulation_instability} Conditions for the Modulational Instability}
%%%%%%%%%%%%%%%%%%%%%%%%%%%%%%%%%%%%%%%%%%%%%%%%%%%%%%%%%%%%%%%%%%%%%%

The nonlinear dispersion relation \cite{dalui2017modulational} of the modulated wave is derived from the above NLSE (\ref{NLSE}) as follows
%Now from the NLSE (\ref{NLSE}), we can find the following nonlinear dispersion relation \cite{dalui2017modulational} of the modulated DA wave  
\begin{eqnarray}\label{modulated_dispersion_relation}
    \Omega^{2} = [P_{d} K^{2}]^{2} \Big( 1 - \frac{2 Q_{d} |\phi_{0}|^{2}}{P_{d} K^{2}} \Big) ,
\end{eqnarray}
where $\Omega$ and $K$ are the modulated wave frequency and modulated wave number respectively.

From the above nonlinear DR (\ref{modulated_dispersion_relation}) of modulated wave, we have derived the following instability conditions:
(i) when $P_{d}Q_{d}<0$ then $\Omega^{2} > 0$ so, the modulated DA wave is stable,  
(ii) when $P_{d}Q_{d}>0$ and  $K \geq K_{c}$ then $\Omega^{2} \geq 0$ so, the modulated DA wave is stable and
(iii) when $P_{d}Q_{d}>0$ and $K < K_{c}$ then $\Omega^{2} < 0$ so, the modulated DA wave is unstable, where $K_{c}=\sqrt{\frac{2Q_{d}|\phi_{0}|^{2}}{P_{d}}}$.
%, i.e., the modulated DA wave is stable or unstable for $K \geq K_{c}$ or  $K < K_{c}$. 

Therefore, for $P_{d}Q_{d}>0$ and $K < K_{c}$, the modulated DA wave is unstable and consequently the modulational growth rate of instability $G$ ($=Im(\Omega)$) is given by the following equation:
\begin{eqnarray}\label{growth_rate_of_instability}
    G^{2} = [P_{d} K^{2}]^{2} \Big( \frac{2 Q_{d} |\phi_{0}|^{2}}{P_{d} K^{2}}-1 \Big).
\end{eqnarray}

For fixed values of $P_{d}$ and $Q_{d}$, the modulational growth rate of instability ($G$) attains its maximum value $G_{max}$ at $ K= \frac{K_{c}}{\sqrt{2}} =\sqrt{\frac{Q_{d}|\phi_{0}|^{2}}{P_{d}}}$, and consequently, the MMGRI $G_{max}$ is given by
\begin{eqnarray}\label{form_of_Gamma_max}
  G_{max} = |Q_{d}||\phi_{0}|^{2}. 
\end{eqnarray}

For $P_{d}Q_{d}>0$, the soliton solution of the NLSE (\ref{NLSE}) is called bright envelop soliton \cite{fedele2002solitary,fedele2002envelope,sikdar2018electrostatic} whereas for $P_{d}Q_{d}<0$ the soliton solution of the NLSE (\ref{NLSE}) is called dark (black and gray) envelop soliton \cite{fedele2002solitary,fedele2002envelope,sikdar2018electrostatic}.

%%%%%%%%%%%%%%%%%%%%%%%%%%%%%%%%%%%%%%%%%%%%%%%%%%%%%%%%%%%%%%%%%%%%%%
\section{\label{Results} Results and discussions}
\begin{figure}[ht]
\begin{center}
\includegraphics[width=80mm]{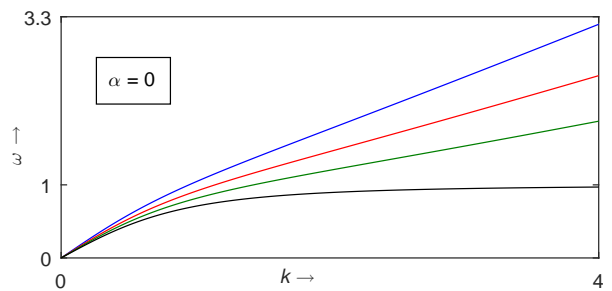}
  \caption{\label{omega_vs_k_alpha_eq_0} The normalized wave frequency ($\omega$) is plotted against the normalized wave number ($k$) for different values of number densities of ions by considering weakly coupled limit ($R \to 0$) of the dusty plasma. Here, blue curve corresponds to $\overline{N}_{i0}=1.1081$ and $\sigma_{d}=0.58$, red curve corresponds to $\overline{N}_{i0}=4.0540$ and $\sigma_{d}=0.33$, green curve corresponds to $\overline{N}_{i0}=6.7567$ and $\sigma_{d}=0.16$, and black curve corresponds to $\overline{N}_{i0}=6.7567$ and $\sigma_{d}=0$. The other values of the parameters are $ \overline{N}_{e0}=\overline{N}_{i0} -1 $, $\sigma_{ie}=0.01$ and $\gamma=1$. % , $\kappa=1$, $\mu = 1$
}
\end{center}
\end{figure}

\begin{figure}[ht]
\begin{center}
\includegraphics[width=80mm]{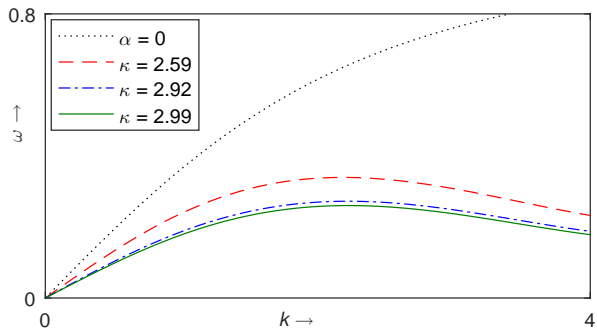}
  \caption{\label{omega_vs_k_diff_kappa_with_alpha} The normalized wave frequency ($\omega$) is plotted against the normalized wave number ($k$) for all set of values of table \ref{Table_2} by considering the strongly coupled limit ($\alpha \neq 0$) of dusty plasma within the QLCA framework. Here, red dashed, blue dash-dotted and green solid curves correspond to the values of Set A, Set B and Set C respectively. The black dotted curve plotted when $\alpha=0$, i.e., when the dusty plasma system is weakly coupled.}
\end{center}
\end{figure}

%%%%%%%%%%%%%%%%%%%%%%%%%%%%%%%%%%%%%%%%%%%%%%%%%%%%%%%%%%%%%%%%%%%%%%
%\subsection{ A Comparison with the experimental results}
First, in order to compare our results with the $T^{\rm eff}$ model \cite{sultana2020dust}, where a more indirect inclusion of strong coupling is done by letting an effective dust temperature represent the strong coupling effect,
%experimental \cite{rosenberg2008note, nosenko2010measurements} study, 
we consider their set of parametric values \cite{sultana2020dust} (given in Table 1) to obtain the linear dispersion relation. The linear dispersion relation with three set of parameters in the weakly coupled limit ($D_{L}$ = 0) is plotted in figure \ref{omega_vs_k_alpha_eq_0} which shows that $\omega \propto k$ trend at relatively higher $k$ value. This same trend has also been recovered from Ref. \cite{sultana2020dust} but in strong coupling limit. It can be concluded that  
our weak coupled ($D_{L}$ = 0) linear dispersion relation shows correspondence with the strong coupled dispersion relation obtained in Ref.~\cite{sultana2020dust}. The linear dispersion relation for these set of parameters, with the QLCA effects ($D_{L}$ $\neq$ 0), is plotted in figure \ref{omega_vs_k_diff_kappa_with_alpha}, which shows the signature of negative dispersion relation. While the negative dispersion is particular manifestation of the strong coupling effects, the $T^{\rm eff}$-model does not predicts this character in the linear dispersion relation.

\begin{table}[ht]
\begin{tabular}{|c c c c |}
\hline
Parameter & Set A & Set B & Set C \\
\hline
$\overline{N}_{i0}$     & $ 1.1081 $              & $ 2.9412 $              & $ 1.9231 $ \\
$\overline{N}_{e0} $    & $ 0.1081 $              &  $ 1.9412 $             & $ 0.9231 $  \\
$\sigma_{ie} $          & $ 0.01 $                & $ 0.02 $                & $ 0.0083 $  \\
$\sigma_{d} $           & $ 2.70 \times 10^{-4} $ & $ 3.53 \times 10^{-4} $ & $ 4.62 \times 10^{-4} $  \\
$\kappa   $             & $ 2.59 $                & $ 2.92 $                & $ 2.99 $                 \\
\hline
\end{tabular}
\caption{\label{Table_2} The table of basic dimensionless parameters involved in the present Yukawa system.}
\end{table}

%----- FIG-1 --------------
%The linear DR (\ref{conventional_DA_LDR}) of DA waves for weakly coupled limit ($R \to 0$, i.e., $D_{L}(k) \to 0$) of dusty plasma is presented in Fig. \ref{omega_vs_k_alpha_eq_0}.
%% for different set of values as mentioned in figure caption where the values of ion number densities taken from Fig. 2 of Ref. \cite{sultana2020dust}.
%% (see Fig. ), i.e., for different values of $\overline{N}_{i0}$. 
%Although Fig. \ref{omega_vs_k_alpha_eq_0} of this problem is plotted for weakly coupled limit (taking $R \to 0$) of the dusty plasma, the qualitative behaviour of the linear DR (\ref{conventional_DA_LDR}) matches with the linear DR of the strongly coupled limit (via effective temperature $T^{\rm eff}$) of dusty plasma (see figure 2 of Ref. \cite{sultana2020dust}) as described in a theoretical work \cite{sultana2020dust}. 

%----- FIG-2 --------------
%For the case of a strongly coupled limit ($\alpha \neq 0$) of dusty plasma within the QLCA framework instead of strong coupling via $T^{\rm eff}$ model  \cite{sultana2020dust} of dusty plasma is investigated. Therefore, in Fig. \ref{omega_vs_k_diff_kappa_with_alpha}, the normalized wave frequency ($\omega$) is plotted against normalized wave number ($k$) for all set of values of table \ref{Table_2}.
%We observed that the normalized wave frequency of strongly coupled dusty plasma decreases with increasing values of the screening parameter ($\kappa$). This results also follow the results Refs. \cite{rosenberg1997dust, khrapak2017practical}.
%%of Rosenberg and Kalman \cite{rosenberg1997dust} (see Fig. 1 of Ref. \cite{rosenberg1997dust}).  
%%From here it can be seen that the second effect of strong %dust-dust correlations is also to reduce the effective dust %plasma frequency

%----- FIG-3 --------------
\begin{figure}[ht]
\begin{center}
\includegraphics[width=80mm]{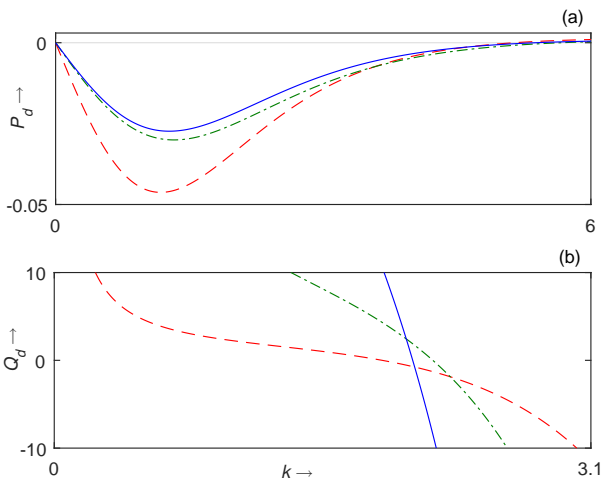}
  \caption{\label{P_and_Q_reproduce} $P_{d}$ and $Q_{d}$ are plotted against $k$ in (a) and (b) respectively for weakly coupled limit of the dusty plasma, i.e., $R \to 0$. Here, the red dashed curve correspond to parametric values $h_{1} = 1.1092 $, $h_{2} = -0.5540 $, $h_{3} = 0.1847$, $\kappa=2.59$, $\sigma_{d}=0.58$,
	%$\overline{N}_{i0}=1.1081$, $\overline{N}_{e0}=0.1081$, $\sigma_{ie}=0.01$,  
	green curve dash dotted curve correspond to parametric values $h_{1} = 2.9800 $, $h_{2} = -1.4702$, $h_{3} = 0.4902$, $\kappa=2.92$, $\sigma_{d}=0.57$,
	%$\overline{N}_{i0}=1.9231$, $\overline{N}_{e0}=0.9231$, $\sigma_{ie}=0.0083$,  
	and blue solid curve correspond to parametric values $h_{1} = 1.9308$, $h_{2} = -0.9615$, $h_{3} = 0.3205$, $\kappa=2.99$, $\sigma_{d}=0.57$.
	%$\overline{N}_{i0}=2.9412$, $\overline{N}_{e0}=1.9412$, $\sigma_{ie}=0.02$, 
	}
\end{center}
\end{figure}
%The $P_{d}$ and $Q_{d}$ are plotted against $k$, for a weakly coupled limit ($D_{L}$ = 0), in figure \ref{P_and_Q_reproduce}, with three set of parameter given in table \ref{Table_2}. 

The dispersive coefficient ($P_{d}$) and nonlinear coefficient ($Q_{d}$) are plotted against $k$, for the weakly coupled limit ($D_{L}$ = 0) of dusty plasma, in figure \ref{P_and_Q_reproduce}, with three set of parameter given in table \ref{Table_2}. The same qualitative nature of $P_{d}$ and $Q_{d}$ was also reported in strongly coupled limit of dusty plasma \cite{sultana2020dust}.

%By considering the weakly coupled limit ($R \to 0$) of the dusty plasma, in figure \ref{P_and_Q_reproduce}, (a) the coefficient of the dispersion term ($P_{d}$) and (b) the coefficient of the nonlinear term ($Q_{d}$) of the NLSE (\ref{NLSE}) are plotted against the normalized wave number ($k$). From figure \ref{P_and_Q_reproduce}, we see that $P_{d}$ and $Q_{d}$ both changes sign within a certain limit of the wave number therefore, we can find the regions $P_{d}Q_{d}<0$ and $P_{d}Q_{d}>0$, i.e., modulational instability regions may exist in the weakly coupled limit of the dusty plasma. The qualitative behavior of our results in weakly coupled limit of the dusty plasma coincide with the results of Ref. \cite{sultana2020dust} in strongly coupled (via $T^{\rm eff}$) limit of the dusty plasma (see Fig. 3 of Ref. \cite{sultana2020dust}).

% for a different set of values. 
% Here, the red dashed curve correspond to parametric values $h_{1} = 1.1092 $, $h_{2} = -0.5540 $, $h_{3} = 0.1847$, $\kappa=2.59$, $\sigma_{d}=0.58$,
 %green curve dash dotted curve correspond to parametric values $h_{1} = 1.9308$, $h_{2} = -0.9615$, $h_{3} = 0.3205$, $\kappa=2.99$, $\sigma_{d}=0.57$,
 %and blue solid curve correspond to parametric values $h_{1} = 2.9800 $, $h_{2} = -1.4702$, $h_{3} = 0.4902$, $\kappa=2.92$, $\sigma_{d}=0.57$. 
%
%----- FIG-4 --------------
\begin{figure}[ht]
\begin{center}
\includegraphics[width=80mm]{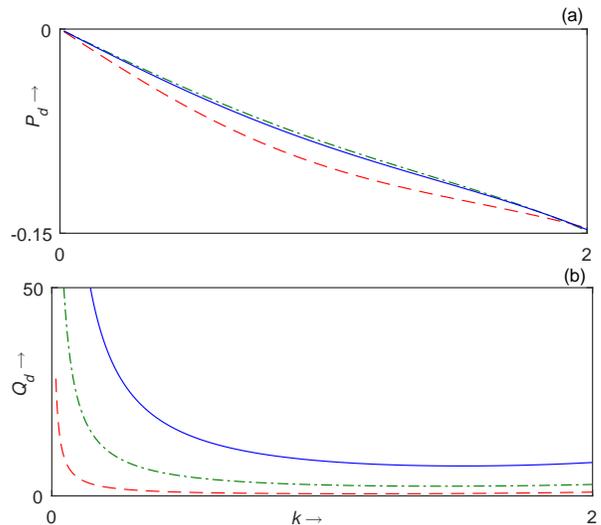}
  \caption{\label{P_and_Q_with_alpha_high_kappa_value} $P_{d}$ and $Q_{d}$ are plotted against $k$ in (a) and (b) respectively for strongly coupled limit of the dusty plasma within the QLCA framework. Here, the red dash dotted curve correspond to parametric values $h_{1} = 1.1092$, $h_{2} = -0.5540$, $h_{3} = 0.1847$, $\kappa=2.59$, $R=1.259$, $\sigma_{d}=0.00027$, green dashed curve correspond to parametric values $h_{1} = 2.9800$, $h_{2} = -1.4702$, $h_{3} = 0.4902$, $\kappa=2.92$, $R=1.292$, $\sigma_{d}=0.000353$ and blue solid curve correspond to parametric values   $h_{1} = 1.9308$, $h_{2} = -0.9615$, $h_{3} = 0.3205$, $\kappa=2.99$, $R=1.299$, $\sigma_{d}=0.000462$.}
\end{center}
\end{figure}
%By considering the strongly coupled limit of the dusty plasma ($\alpha \neq 0$) within the QLCA framework, in figure \ref{P_and_Q_reproduce_with_alpha}, (a) the coefficient of the dispersion term ($P_{d}$) and (b) the coefficient of the nonlinear term ($Q_{d}$) of the NLSE (\ref{NLSE}) are plotted against the normalized wave number ($k$). 

\begin{figure}[ht]
\begin{center}
\includegraphics[width=80mm]{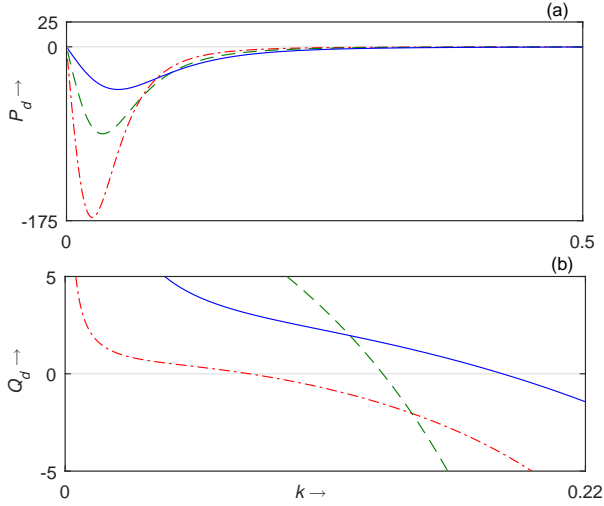}
  \caption{\label{P_and_Q_reproduce_with_alpha} $P_{d}$ and $Q_{d}$ are plotted against $k$ in (a) and (b) respectively for strongly coupled limit of the dusty plasma within QLCA framework. Here, the red dash dotted curve correspond to parametric values $h_{1} = 1.1092$, $h_{2} = -0.5540$, $h_{3} = 0.1847$, $\kappa=0.05$, $R=1.005$, $\sigma_{d}=0.00027$, green dashed curve correspond to parametric values  $h_{1} = 2.9800$, $h_{2} = -1.4702$, $h_{3} = 0.4902$, $\kappa=0.07$, $R=1.007$, $\sigma_{d}=0.000353$, and blue solid curve correspond to parametric values  $h_{1} = 1.9308$, $h_{2} = -0.9615$, $h_{3} = 0.3205$, $\kappa=0.1$, $R=1.01$, $\sigma_{d}=0.000462$.}
\end{center}
\end{figure}

\begin{figure}[ht]
\begin{center}
\includegraphics[width=80mm]{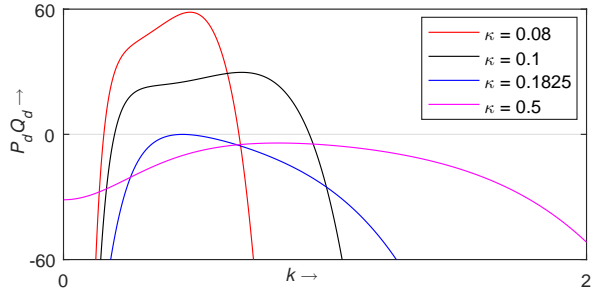}
  \caption{\label{PQ_vs_k_for_diff_kappa_Set_2_9_diff_kappa} For strongly coupled limit of dusty plasma, the product of the coefficients of the nonlinear and dispersive term ($P_{d}Q_{d}$) is plotted against $k$ for different values of the screening parameter $\kappa$ and for $h_{1} = 2.9800$, $h_{2} = -1.4702$, $h_{3} = 0.4902$ $\sigma_{d}=0.000353$.}
\end{center}
\end{figure}

From Fig. \ref{P_and_Q_reproduce} and \ref{P_and_Q_with_alpha_high_kappa_value}, we compare the weak and strong coupling effects on the coefficients $P_{d}$ and $Q_{d}$, as their product ($P_{d}Q_{d}$) decide the stable and unstable region.  
The red dashed, green dash dotted and blue solid line representing the coefficient $P_{d}$ and $Q_{d}$ for parameter $\kappa$ = 2.59, 2.92 and 2.99, respectively, with small value of $\sigma_{d}$. In both weak and strong coupling limit, the dispersive coefficient $P_{d}$ always remain negative, whereas in the weak coupling limit the nonlinear coefficient $Q_{d}$ has both negative as well as positive values depending upon the value of $\kappa$, and in the strongly coupled system it always remains positive at relatively high value of $\kappa $. In order to  explore the lower $\kappa$ effects on the coefficients $P_{d}$ and $Q_{d}$ in the strong coupling limit, we have plotted the $P_{d}$ and $Q_{d}$ against $k$ with different small values of $\kappa$ in figure \ref{P_and_Q_reproduce_with_alpha}.     
 It has also been identified, from Fig. \ref{P_and_Q_with_alpha_high_kappa_value}, a negative threshold values of coefficient $Q_{d}$ arising at different value of wave-vector $k$. This negative threshold value appeared at relatively higher $k$  by increasing the value of $\kappa$ and, after $\kappa$ = 0.1825 coefficient $Q_{d}$ again attains positive value which remains positive for all value of $k$.
  The product $P_{d}Q_{d}$  against $k$ has been plotted in Fig. \ref{PQ_vs_k_for_diff_kappa_Set_2_9_diff_kappa} with different small $\kappa$ values. It has been identified from Fig.  \ref{PQ_vs_k_for_diff_kappa_Set_2_9_diff_kappa} that the unstable region ($P_{d}Q_{d}$ $>$ 0), as presented with red and black curves, increases with an increment of $\kappa$ from 0.08 to 0.1. The  threshold  has been reached at $\kappa$ = 0.1825, as represented by blue line, after that modulated wave is stable ($P_{d}Q_{d}$ $<$ 0) for all $k$ and $\kappa$.
 
%----- FIG-5 --------------
 
%----- FIG-6 --------------

\begin{figure}[ht]
\begin{center}
\includegraphics[width=80mm]{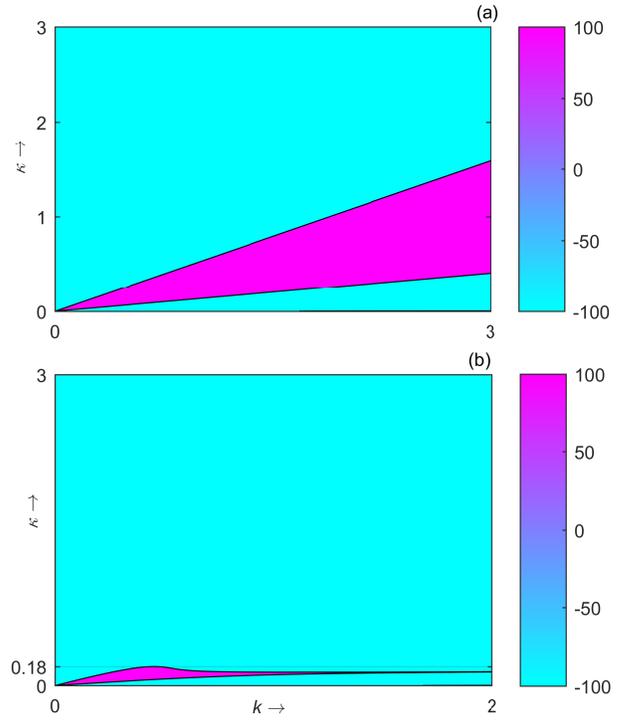}
  \caption{\label{Contour_plot_PQ_kappa_vs_k_Set_B_with_alpha_0_to_3_1000} In Fig. \ref{Contour_plot_PQ_kappa_vs_k_Set_B_with_alpha_0_to_3_1000}(a) and \ref{Contour_plot_PQ_kappa_vs_k_Set_B_with_alpha_0_to_3_1000}(b), the contour of the product of the coefficients of dispersive and nonlinear terms, i.e., $P_{d}Q_{d}$ is represented here with respect to the wave number $k$ against the screening parameter $\kappa$ for the effect of weakly and strongly coupled dusty plasma respectively. For, $h_{1}=2.9800$, $h_{2}=-1.4702$, $h_{3}=0.4902$, $\sigma_{d}=0.000353$ and $\gamma=1$.}
\end{center}
\end{figure}

In order to compare the stable and unstable regions, in more detail, for weakly and strongly coupled limit, the contour plot of the product $P_{d}Q_{d}$ on $k-\kappa$ plane has been plotted in Fig. \ref{Contour_plot_PQ_kappa_vs_k_Set_B_with_alpha_0_to_3_1000}. 
It has been observed that a relatively larger unstable region (pink region) in weakly coupled limit, whereas in strongly coupled limit this unstable region is suppressed to a very small portion. In comparison to analysis of modulational instability in a one dimensional chain \cite{amin1998modulational_dustlattice}, where they have predicted an unstable region \cite{amin1998modulational_dustlattice} for wide  range  of $\kappa$ value, our QLCA based analysis, which incorporates isotropy (3D structure) and explicit localization of constituent particles, has found completely stable region (cyan color) arising beyond $\kappa$ = 0.183 [see Fig \ref{Contour_plot_PQ_kappa_vs_k_Set_B_with_alpha_0_to_3_1000}(b) and Fig. \ref{PQ_vs_k_for_diff_kappa_Set_2_9_diff_kappa}]. It can be seen from Fig. \ref{Contour_plot_PQ_kappa_vs_k_Set_B_with_alpha_0_to_3_1000}(b) that the modulated wave become stable above threshold value $\kappa$ = 0.183 for all $k$.  

\begin{figure}
\begin{center}
\includegraphics[width=80mm]{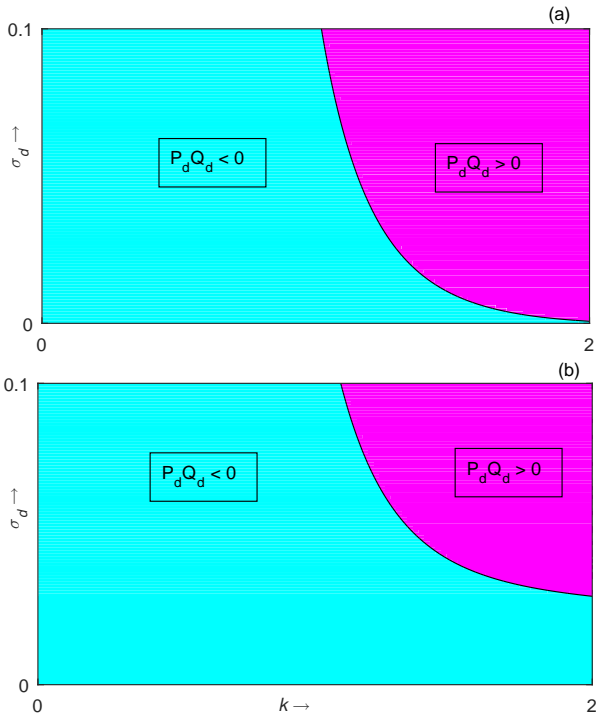}
  \caption{\label{Contour_plot_PQ_sigma_d_vs_k_setB_100} In Fig. \ref{Contour_plot_PQ_sigma_d_vs_k_setB_100}(a) and \ref{Contour_plot_PQ_sigma_d_vs_k_setB_100}(b), the contour of the product of the coefficients of dispersive and nonlinear terms, i.e., $P_{d}Q_{d}$ is represented here with respect to the wave number $k$ against $\sigma_{d}$ for weakly and strongly coupled dusty plasma respectively. Here, $h_{1}=2.9800$, $h_{2}=-1.4702$, $h_{3}=0.4902$, $\kappa=1.1$, $R=1.11$ and $\gamma=1$.}
\end{center}
\end{figure}
The effect of dust temperature via parameter $\sigma_{d}$ on the modulated wave has been presented in Fig \ref{Contour_plot_PQ_sigma_d_vs_k_setB_100}, for both weak and strong coupling limit. For strong coupling limit, the contour plot of $P_{d}Q_{d}$ in $k-\sigma_{d}$ space shows that temperature enhances the unstable region (not presented in Fig.~\ref{Contour_plot_PQ_kappa_vs_k_Set_B_with_alpha_0_to_3_1000}(b)). It can be seen in Fig.~\ref{Contour_plot_PQ_sigma_d_vs_k_setB_100} that unstable region, represented by pink color region ($P_{d}Q_{d}$ $>$ 0), appears at relatively higher $k$ and $\sigma_{d}$ values. 
It can be concluded from these observations that temperature competes with the QLCA effects and at higher temperatures, the thermal effects dominate over the QLCA effects. The same trend has also been observed for weakly coupled limit of the present system.

\begin{figure}[ht]
\begin{center}
\includegraphics[width=80mm]{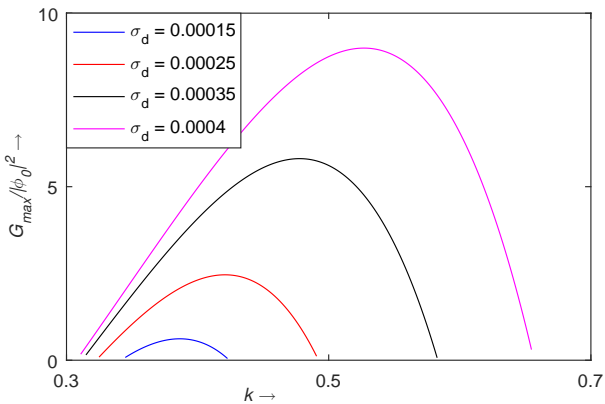}
  \caption{\label{Gamma_max_vs_k_diff_sigma_d} $G_{max}/|\phi_{0}|^{2}$ is plotted against $k$ for different values of $\sigma_{d}$ and for $h_{1}=2.9800$, $h_{2}=-1.4702$, $h_{3}=0.4902$, $\kappa=0.15$, $R= 1.015$ and $\gamma=1$.}
\end{center}
\end{figure}

%In figure \ref{Gamma_max_vs_k_diff_sigma_d}, MMGRI $G_{max}/|\phi_{0}|^{2}$ is plotted against $k$ for different values of $\sigma_{d}$ with the effect of strongly coupled limit dusty plasma and for $h_{1}=2.9800$, $h_{2}=-1.4702$, $h_{3}=0.4902$, $\kappa=0.15$, $R=1+\frac{\kappa}{10} = 1.015$ and $\gamma=1$. Here, blue, red, black and pink color curves correspond to $\sigma_{d}=0.00015$, $\sigma_{d}=0.00025$, $\sigma_{d}=0.00035$ and $\sigma_{d}=0.0004$ respectively. From figure \ref{Gamma_max_vs_k_diff_sigma_d}, we see that the region of existence of the maximum modulational growth rate of instability increases with increasing $\sigma_{d}$. %(or decreasing $Z_{d}$ since $\sigma_{d} = \frac{T_{d}}{Z_{d}T_{i}}$). 
%Also, the MMGRI increases with increasing $\sigma_{d}$. 

For strongly coupled limit, the maximum modulational growth rate of instability ($G_{max}/|\phi_{0}|^{2}$) is plotted against $k$ in Fig. \ref{Gamma_max_vs_k_diff_sigma_d} for different value of dust temperatures via $\sigma_{d}$ within the QLCA framework. The blue, red, black and pink color curves correspond to $\sigma_{d}=0.00015$, $\sigma_{d}=0.00025$, $\sigma_{d}=0.00035$ and $\sigma_{d}=0.0004$, respectively. It has been shown that the region of existence of the maximum modulational growth rate of instability increases with increasing $\sigma_{d}$. We can conclude that the dust temperature enhances the instability in the Yukawa system this trend is also predicted from the figure \ref{Contour_plot_PQ_sigma_d_vs_k_setB_100}(b).
 %(or decreasing $Z_{d}$ since $\sigma_{d} = \frac{T_{d}}{Z_{d}T_{i}}$). 

%Also, the MMGRI increases with increasing $\sigma_{d}$.
%
%
%there exists a critical value $k^{c}$ of the wave number ($k$) such that the MMGRI increases for  maximum modulational growth
%rate of instability increases with the increasing wave
%number.  

%----- FIG-10 --------------
\begin{figure}[ht]
\begin{center}
\includegraphics[width=80mm]{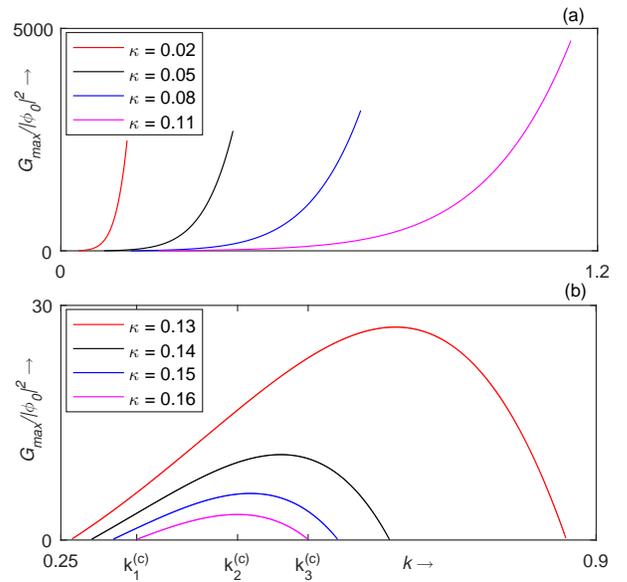}
  \caption{\label{Gamma_max_vs_k_diff_kappa_set_B_100} $G_{max}/|\phi_{0}|^{2}$ is plotted against $k$ for different values of $\kappa$ and for $h_{1}=2.9800$, $h_{2}=-1.4702$, $h_{3}=0.4902$, $\sigma_{d}=0.000353$ and $\gamma=1$.}
\end{center}
\end{figure}

For strongly coupled limit, $G_{max}/|\phi_{0}|^{2}$ is plotted against $k$ for different values of $\kappa$ in Fig \ref{Gamma_max_vs_k_diff_kappa_set_B_100}.
% and the parameters $\sigma_{d}=0.000353$, and $\gamma=1$, are considered to be fix. 
%From figures \ref{Gamma_max_vs_k_diff_kappa_set_B_100}(a) and \ref{Gamma_max_vs_k_diff_kappa_set_B_100}(b), 
It can be seen that from Fig.\ref{Gamma_max_vs_k_diff_kappa_set_B_100}(a), that the MMGRI is increasing with $\kappa$ for 0.02 $\leq \kappa \leq $ 0.11. 
The MMGRI has been plotted with relatively higher values of $\kappa$ in Fig. \ref{Gamma_max_vs_k_diff_kappa_set_B_100}(b), which shows that for fixed value of $\kappa$, the MMGRI first increase and after attaining a maximum value it again decreases until hit to the zero value. 
%The red, black, blue and pink line correspond to $\kappa$ = 0.13, 0.14, 0.15 and 0.16, respectively.
The peak value of MMGRI is reducing with $\kappa$ for 0.13 $\leq \kappa \leq $ 0.16 and as we already predicted that the MMGRI becomes zero at $\kappa$ = 0.183. It can be concluded that the modulated wave will become stable after $\kappa$ = 0.183 for all values of $k$.
For example, we see that there exist the critical values $k_{1}^{(c)} = 0.34201$, $k_{2}^{(c)}= 0.46458$ and $k_{3}^{(c)}=0.55026$ of $k$ such that the MMGRI increases with increasing wave number ($k$) for $k_{1}^{(c)} < k < k_{2}^{(c)}$ and the MMGRI decreases with increasing wave number ($k$) for $k_{2}^{(c)} < k < k_{3}^{(c)}$ in figure \ref{Gamma_max_vs_k_diff_kappa_set_B_100}(b). Finally, we can conclude that there exist the critical value $\kappa_{1}^{(c)} = 0.1284$ and $\kappa_{2}^{(c)} = 0.1821$ of $\kappa$ such that the MMGRI exists for $0< \kappa < \kappa_{2}^{(c)}$ and the MMGRI does not exists for $ \kappa_{2}^{(c)} < \kappa < 3$ because for these values of $\kappa$ the modulated wave became stable. This fact also confirms from  Fig. \ref{Contour_plot_PQ_kappa_vs_k_Set_B_with_alpha_0_to_3_1000}(b).

Before concluding the discussion, a relevance can be drawn between dusty 
plasma excitation treated here and, for example, with the observation in 
RF field trapping of the ultracold ions \cite{zhou2021mass} where motion of 
signaling ion species was found to be tunable at the edge of the stability region as a result of ions 
being quasi-localized. 
The conclusion that collective ion interaction remains responsible for
the observed delocalization in the boundary zone does indicate the role 
of constructively interacting collective ion excitation. The role 
of temperature of trapped species in this case can indeed be expected to 
be marginal as for Mathieu parameter $q\sim 1$, the mechanical motion of 
ions is entirely attributed to the collective (resonant) effect.

%%%%%%%%%%%%%%%%%%%%%%%%%%%%%%%%%%%%%%%%%%%%%%%%%%%%%%%%%%%%%%%%%%%%%%
\section{\label{Conclusions} Conclusions}

In this article, the QLCA based model has been adopted to study the MI of the DA waves in a strongly coupled Yukawa system consisting of negatively charged dust grains embedded in a polarizable plasma medium following the Boltzmann distribution.  In order to study the modulated wave, we have derived the NLSE (\ref{NLSE}) using RPM \cite{taniuti1969perturbation,asano1969perturbation}. It has been seen from the linear analysis that the DA wave frequency is reduced when the strong coupling effects are incorporated via QLCA framework \cite{rosenberg1997dust}. 
%For strongly coupled dusty plasma we have seen that both the phase velocity ($\frac{\omega}{k}$) and group velocity ($V_{g}$) decreases with respect to the wave number ($k$).
For the weak coupling case, it has been observed that the qualitative behaviour of the linear dispersion relation matches with the strongly coupled limit ($T^{\rm eff}$ model) \cite{sultana2020dust} in the dusty plasma. 
The MI of DA waves is numerically investigated for both the cases viz., for weakly and strongly coupled limits of the dusty plasma. It has been observed that in weakly coupled limit a relatively larger unstable region is recovered, whereas in strongly coupled limit this region is reduced to a very small zone of the parameter space. In comparison to analysis of modulation instability in a one-dimensional chain \cite{amin1998modulational_dustlattice}, where existing studies have predicted an unstable region \cite{amin1998modulational_dustlattice} for wide range of $\kappa$ value, our QLCA based analysis incorporating explicit and isotropic localization of constituent particles, has recovered unstable region upto a relatively smaller value of $\kappa$ = 0.183 for a typical (small) dust temperature value $T_{d} \sim 10^{-4} Z_{d}T_{i}$.  
For strong coupling limit, the contour plot of $P_{d}Q_{d}$ in $k-\sigma_{d}$ space shows that the larger dust temperature enhances the unstable region dimension in the parameter space.
 %The peak value of MMGRI is reducing with $\kappa$ in the range from 0.13 to 0.16 and that the MMGRI become zero at $\kappa$ = 0.183. 
 The peak value of MMGRI is reducing with $\kappa$ for 0.13 $\leq \kappa \leq$ 0.16 and that MMGRI become zero at $\kappa$ = 0.183. 
%To maintain extreme state of a system, such as perfect lattice ($\Gamma$ $>>$ 1), and weakly coupled state ($\Gamma$ $<<$ 1) requires extreme values of their parameters (temperature and density), whereas the quasi-crystal state is naturally available in the physical systems and has a potential to attain either of these two extreme states. 
The analysis on the instability criteria of a modulated wave, presented here, is largely applicable to quasi-crystalline state (amorphous solid) in which both free motion as well as localization of the constituent particles coexist. The presented results are therefore expected to cover a wide range of natural systems where modulational instability is the prime mechanism for the weakly nonlinear collective effects. As a relevant example, the case of collective interaction driven delocalization of RF trapped ultracold ions is discussed which is observed at the stability boundary in a recent experiment where the background 
interference of the RF trapping field drops sharply, leaving the trapped ion species to be in a quasi-localized state.
%In order to explore this intermediate state (liquid state), we adopt a more sophisticated theoretical model , namely Quasi-localized charged approximation (QLCA), which has been proven to more reliable predictor of such state.
% We have seen that for the introduction of strong coupling in the system the stable region ($P_{d}Q_{d} < 0$) increases and unstable region ($P_{d}Q_{d} > 0$) decreases.
%The MMGRI theoretically and numerically investigated for strongly coupled limit of dusty plasma. We have seen that the MMGRI as well as the region of existence of MMGRI increases with increasing $\sigma_{d}$. 
%
%The MMGRI exists for $0< \kappa < \kappa_{2}^{(c)}$. And the MMGRI does not exists for $ \kappa_{2}^{(c)} < \kappa < 3$ because the product of the coefficients of dispersive and nonlinear terms of NLSE is less than $0$, i.e., $P_{d}Q_{d} < 0$. So, for any values of $\kappa \in (\kappa_{2}^{(c)}, ~ 3)$ the modulated DA wave is stable. 
%Finally, we have shown the existence of bright and dark envelope solitary waves in strongly coupled dusty plasma.
Within the QLCA framework, the nonlinear excitations of MI of DA waves in strongly coupled dusty plasma can be a treated in presence of a magnetic field as a future study. The investigation on existence of envelope solitary waves, analytically as well as numerically, in a strongly coupled Yukawa system within QLCA framework can be another area to be explored.

%In future, we want to investigate the existence of envelope solitary waves analytically and numerically in strongly coupled Yukawa system. 

\bibliographystyle{apsrev4-1}
\bibliography{paper_2}

\begin{thebibliography}{66}%
\makeatletter
\providecommand \@ifxundefined [1]{%
 \@ifx{#1\undefined}
}%
\providecommand \@ifnum [1]{%
 \ifnum #1\expandafter \@firstoftwo
 \else \expandafter \@secondoftwo
 \fi
}%
\providecommand \@ifx [1]{%
 \ifx #1\expandafter \@firstoftwo
 \else \expandafter \@secondoftwo
 \fi
}%
\providecommand \natexlab [1]{#1}%
\providecommand \enquote  [1]{``#1''}%
\providecommand \bibnamefont  [1]{#1}%
\providecommand \bibfnamefont [1]{#1}%
\providecommand \citenamefont [1]{#1}%
\providecommand \href@noop [0]{\@secondoftwo}%
\providecommand \href [0]{\begingroup \@sanitize@url \@href}%
\providecommand \@href[1]{\@@startlink{#1}\@@href}%
\providecommand \@@href[1]{\endgroup#1\@@endlink}%
\providecommand \@sanitize@url [0]{\catcode `\\12\catcode `\$12\catcode
  `\&12\catcode `\#12\catcode `\^12\catcode `\_12\catcode `\%12\relax}%
\providecommand \@@startlink[1]{}%
\providecommand \@@endlink[0]{}%
\providecommand \url  [0]{\begingroup\@sanitize@url \@url }%
\providecommand \@url [1]{\endgroup\@href {#1}{\urlprefix }}%
\providecommand \urlprefix  [0]{URL }%
\providecommand \Eprint [0]{\href }%
\providecommand \doibase [0]{http://dx.doi.org/}%
\providecommand \selectlanguage [0]{\@gobble}%
\providecommand \bibinfo  [0]{\@secondoftwo}%
\providecommand \bibfield  [0]{\@secondoftwo}%
\providecommand \translation [1]{[#1]}%
\providecommand \BibitemOpen [0]{}%
\providecommand \bibitemStop [0]{}%
\providecommand \bibitemNoStop [0]{.\EOS\space}%
\providecommand \EOS [0]{\spacefactor3000\relax}%
\providecommand \BibitemShut  [1]{\csname bibitem#1\endcsname}%
\let\auto@bib@innerbib\@empty
%</preamble>
\bibitem [{\citenamefont {Koester}\ and\ \citenamefont
  {Sch{\"o}nberner}(1986)}]{koester1986evolution}%
  \BibitemOpen
  \bibfield  {author} {\bibinfo {author} {\bibfnamefont {D.}~\bibnamefont
  {Koester}}\ and\ \bibinfo {author} {\bibfnamefont {D.}~\bibnamefont
  {Sch{\"o}nberner}},\ }\href@noop {} {\bibfield  {journal} {\bibinfo
  {journal} {Astron. Astrophys.}\ }\textbf {\bibinfo {volume} {154}},\ \bibinfo
  {pages} {125} (\bibinfo {year} {1986})}\BibitemShut {NoStop}%
\bibitem [{\citenamefont {Kouveliotou}\ \emph {et~al.}(2001)\citenamefont
  {Kouveliotou}, \citenamefont {Ventura}, \citenamefont {van~den Heuvel},\ and\
  \citenamefont {van~den Heuvel}}]{kouveliotou2001neutron}%
  \BibitemOpen
  \bibfield  {author} {\bibinfo {author} {\bibfnamefont {C.}~\bibnamefont
  {Kouveliotou}}, \bibinfo {author} {\bibfnamefont {J.~E.}\ \bibnamefont
  {Ventura}}, \bibinfo {author} {\bibfnamefont {E.~P.}\ \bibnamefont {van~den
  Heuvel}}, \ and\ \bibinfo {author} {\bibfnamefont {E.~P.~J.}\ \bibnamefont
  {van~den Heuvel}},\ }\href@noop {} {\emph {\bibinfo {title} {The Neutron
  Star: Black Hole Connection}}},\ Vol.\ \bibinfo {volume} {567}\ (\bibinfo
  {publisher} {Springer Science \& Business Media},\ \bibinfo {year}
  {2001})\BibitemShut {NoStop}%
\bibitem [{\citenamefont {Chabrier}\ \emph {et~al.}(2002)\citenamefont
  {Chabrier}, \citenamefont {Douchin},\ and\ \citenamefont
  {Potekhin}}]{chabrier2002dense}%
  \BibitemOpen
  \bibfield  {author} {\bibinfo {author} {\bibfnamefont {G.}~\bibnamefont
  {Chabrier}}, \bibinfo {author} {\bibfnamefont {F.}~\bibnamefont {Douchin}}, \
  and\ \bibinfo {author} {\bibfnamefont {A.~Y.}\ \bibnamefont {Potekhin}},\
  }\href@noop {} {\bibfield  {journal} {\bibinfo  {journal} {J. Phys.: Condens.
  Matter}\ }\textbf {\bibinfo {volume} {14}},\ \bibinfo {pages} {9133}
  (\bibinfo {year} {2002})}\BibitemShut {NoStop}%
\bibitem [{\citenamefont {Shukla}\ \emph {et~al.}(1996)\citenamefont {Shukla},
  \citenamefont {Vladimirov},\ and\ \citenamefont
  {Nambu}}]{shukla1996stimulated}%
  \BibitemOpen
  \bibfield  {author} {\bibinfo {author} {\bibfnamefont {P.~K.}\ \bibnamefont
  {Shukla}}, \bibinfo {author} {\bibfnamefont {S.~V.}\ \bibnamefont
  {Vladimirov}}, \ and\ \bibinfo {author} {\bibfnamefont {M.}~\bibnamefont
  {Nambu}},\ }\href@noop {} {\bibfield  {journal} {\bibinfo  {journal} {Phys.
  Scr.}\ }\textbf {\bibinfo {volume} {53}},\ \bibinfo {pages} {89} (\bibinfo
  {year} {1996})}\BibitemShut {NoStop}%
\bibitem [{\citenamefont {Rosenberg}\ and\ \citenamefont
  {Shukla}(2011)}]{rosenberg2011instabilities}%
  \BibitemOpen
  \bibfield  {author} {\bibinfo {author} {\bibfnamefont {M.}~\bibnamefont
  {Rosenberg}}\ and\ \bibinfo {author} {\bibfnamefont {P.~K.}\ \bibnamefont
  {Shukla}},\ }\href@noop {} {\bibfield  {journal} {\bibinfo  {journal} {Phys.
  Scr.}\ }\textbf {\bibinfo {volume} {83}},\ \bibinfo {pages} {015503}
  (\bibinfo {year} {2011})}\BibitemShut {NoStop}%
\bibitem [{\citenamefont {Golden}\ \emph {et~al.}(1992)\citenamefont {Golden},
  \citenamefont {Kalman},\ and\ \citenamefont {Wyns}}]{golden1992dielectric2d}%
  \BibitemOpen
  \bibfield  {author} {\bibinfo {author} {\bibfnamefont {K.~I.}\ \bibnamefont
  {Golden}}, \bibinfo {author} {\bibfnamefont {G.}~\bibnamefont {Kalman}}, \
  and\ \bibinfo {author} {\bibfnamefont {P.}~\bibnamefont {Wyns}},\ }\href@noop
  {} {\bibfield  {journal} {\bibinfo  {journal} {Phys. Rev. A}\ }\textbf
  {\bibinfo {volume} {46}},\ \bibinfo {pages} {3463} (\bibinfo {year}
  {1992})}\BibitemShut {NoStop}%
\bibitem [{\citenamefont {Kalman}\ \emph {et~al.}(1999)\citenamefont {Kalman},
  \citenamefont {Valtchinov},\ and\ \citenamefont
  {Golden}}]{kalman1999collective}%
  \BibitemOpen
  \bibfield  {author} {\bibinfo {author} {\bibfnamefont {G.}~\bibnamefont
  {Kalman}}, \bibinfo {author} {\bibfnamefont {V.}~\bibnamefont {Valtchinov}},
  \ and\ \bibinfo {author} {\bibfnamefont {K.~I.}\ \bibnamefont {Golden}},\
  }\href@noop {} {\bibfield  {journal} {\bibinfo  {journal} {Phys. Rev. Lett.}\
  }\textbf {\bibinfo {volume} {82}},\ \bibinfo {pages} {3124} (\bibinfo {year}
  {1999})}\BibitemShut {NoStop}%
\bibitem [{\citenamefont {Rosenberg}\ and\ \citenamefont
  {Kalman}(1998)}]{rosenberg1998effect}%
  \BibitemOpen
  \bibfield  {author} {\bibinfo {author} {\bibfnamefont {M.}~\bibnamefont
  {Rosenberg}}\ and\ \bibinfo {author} {\bibfnamefont {G.}~\bibnamefont
  {Kalman}},\ }\href@noop {} {\bibfield  {journal} {\bibinfo  {journal} {AIP
  Conference Proceedings}\ }\textbf {\bibinfo {volume} {446}},\ \bibinfo
  {pages} {135} (\bibinfo {year} {1998})}\BibitemShut {NoStop}%
\bibitem [{\citenamefont {Merlino}\ \emph {et~al.}(1998)\citenamefont
  {Merlino}, \citenamefont {Barkan}, \citenamefont {Thompson},\ and\
  \citenamefont {D’angelo}}]{merlino1998laboratory}%
  \BibitemOpen
  \bibfield  {author} {\bibinfo {author} {\bibfnamefont {R.~L.}\ \bibnamefont
  {Merlino}}, \bibinfo {author} {\bibfnamefont {A.}~\bibnamefont {Barkan}},
  \bibinfo {author} {\bibfnamefont {C.}~\bibnamefont {Thompson}}, \ and\
  \bibinfo {author} {\bibfnamefont {N.}~\bibnamefont {D’angelo}},\
  }\href@noop {} {\bibfield  {journal} {\bibinfo  {journal} {Phys. Plasmas}\
  }\textbf {\bibinfo {volume} {5}},\ \bibinfo {pages} {1607} (\bibinfo {year}
  {1998})}\BibitemShut {NoStop}%
\bibitem [{\citenamefont {Fortov}\ \emph {et~al.}(2005)\citenamefont {Fortov},
  \citenamefont {Ivlev}, \citenamefont {Khrapak}, \citenamefont {Khrapak},\
  and\ \citenamefont {Morfill}}]{fortov2005complex}%
  \BibitemOpen
  \bibfield  {author} {\bibinfo {author} {\bibfnamefont {V.~E.}\ \bibnamefont
  {Fortov}}, \bibinfo {author} {\bibfnamefont {A.~V.}\ \bibnamefont {Ivlev}},
  \bibinfo {author} {\bibfnamefont {S.~A.}\ \bibnamefont {Khrapak}}, \bibinfo
  {author} {\bibfnamefont {A.~G.}\ \bibnamefont {Khrapak}}, \ and\ \bibinfo
  {author} {\bibfnamefont {G.~E.}\ \bibnamefont {Morfill}},\ }\href@noop {}
  {\bibfield  {journal} {\bibinfo  {journal} {Phys. Rep.}\ }\textbf {\bibinfo
  {volume} {421}},\ \bibinfo {pages} {1} (\bibinfo {year} {2005})}\BibitemShut
  {NoStop}%
\bibitem [{\citenamefont {Shukla}\ and\ \citenamefont
  {Mamun}(2015)}]{shukla2015introduction}%
  \BibitemOpen
  \bibfield  {author} {\bibinfo {author} {\bibfnamefont {P.~K.}\ \bibnamefont
  {Shukla}}\ and\ \bibinfo {author} {\bibfnamefont {A.~A.}\ \bibnamefont
  {Mamun}},\ }\href@noop {} {\emph {\bibinfo {title} {Introduction to dusty
  plasma physics}}}\ (\bibinfo  {publisher} {CRC press},\ \bibinfo {year}
  {2015})\BibitemShut {NoStop}%
\bibitem [{\citenamefont {Horanyi}\ and\ \citenamefont
  {Mendis}(1985)}]{horanyi1985trajectories}%
  \BibitemOpen
  \bibfield  {author} {\bibinfo {author} {\bibfnamefont {M.}~\bibnamefont
  {Horanyi}}\ and\ \bibinfo {author} {\bibfnamefont {D.~A.}\ \bibnamefont
  {Mendis}},\ }\href@noop {} {\bibfield  {journal} {\bibinfo  {journal}
  {Astrophys. J.}\ }\textbf {\bibinfo {volume} {294}},\ \bibinfo {pages} {357}
  (\bibinfo {year} {1985})}\BibitemShut {NoStop}%
\bibitem [{\citenamefont {Horanyi}\ and\ \citenamefont
  {Mendis}(1986)}]{horanyi1986effects}%
  \BibitemOpen
  \bibfield  {author} {\bibinfo {author} {\bibfnamefont {M.}~\bibnamefont
  {Horanyi}}\ and\ \bibinfo {author} {\bibfnamefont {D.~A.}\ \bibnamefont
  {Mendis}},\ }\href@noop {} {\bibfield  {journal} {\bibinfo  {journal} {The
  Astrophysical Journal}\ }\textbf {\bibinfo {volume} {307}},\ \bibinfo {pages}
  {800} (\bibinfo {year} {1986})}\BibitemShut {NoStop}%
\bibitem [{\citenamefont {Goertz}(1989)}]{goertz1989dusty}%
  \BibitemOpen
  \bibfield  {author} {\bibinfo {author} {\bibfnamefont {C.~K.}\ \bibnamefont
  {Goertz}},\ }\href@noop {} {\bibfield  {journal} {\bibinfo  {journal} {Rev.
  Geophys.}\ }\textbf {\bibinfo {volume} {27}},\ \bibinfo {pages} {271}
  (\bibinfo {year} {1989})}\BibitemShut {NoStop}%
\bibitem [{\citenamefont {Northrop}(1992)}]{northrop1992dusty}%
  \BibitemOpen
  \bibfield  {author} {\bibinfo {author} {\bibfnamefont {T.~G.}\ \bibnamefont
  {Northrop}},\ }\href@noop {} {\bibfield  {journal} {\bibinfo  {journal}
  {Phys. Scr.}\ }\textbf {\bibinfo {volume} {45}},\ \bibinfo {pages} {475}
  (\bibinfo {year} {1992})}\BibitemShut {NoStop}%
\bibitem [{\citenamefont {Tsytovich}(1997)}]{tsytovich1997dust}%
  \BibitemOpen
  \bibfield  {author} {\bibinfo {author} {\bibfnamefont {V.~N.}\ \bibnamefont
  {Tsytovich}},\ }\href@noop {} {\bibfield  {journal} {\bibinfo  {journal}
  {Phys. Uspekhi}\ }\textbf {\bibinfo {volume} {40}},\ \bibinfo {pages} {53}
  (\bibinfo {year} {1997})}\BibitemShut {NoStop}%
\bibitem [{\citenamefont {Whipple}(1981)}]{whipple1981potentials}%
  \BibitemOpen
  \bibfield  {author} {\bibinfo {author} {\bibfnamefont {E.~C.}\ \bibnamefont
  {Whipple}},\ }\href@noop {} {\bibfield  {journal} {\bibinfo  {journal} {Rep.
  Prog. Phys.}\ }\textbf {\bibinfo {volume} {44}},\ \bibinfo {pages} {1197}
  (\bibinfo {year} {1981})}\BibitemShut {NoStop}%
\bibitem [{\citenamefont {Robinson}\ and\ \citenamefont
  {Coakley}(1992)}]{robinson1992spacecraft}%
  \BibitemOpen
  \bibfield  {author} {\bibinfo {author} {\bibfnamefont {P.~A.}\ \bibnamefont
  {Robinson}}\ and\ \bibinfo {author} {\bibfnamefont {P.}~\bibnamefont
  {Coakley}},\ }\href@noop {} {\bibfield  {journal} {\bibinfo  {journal} {IEEE
  Trans. Electr. Insul.}\ }\textbf {\bibinfo {volume} {27}},\ \bibinfo {pages}
  {944} (\bibinfo {year} {1992})}\BibitemShut {NoStop}%
\bibitem [{\citenamefont {Rosenberg}\ and\ \citenamefont
  {Kalman}(1997)}]{rosenberg1997dust}%
  \BibitemOpen
  \bibfield  {author} {\bibinfo {author} {\bibfnamefont {M.}~\bibnamefont
  {Rosenberg}}\ and\ \bibinfo {author} {\bibfnamefont {G.}~\bibnamefont
  {Kalman}},\ }\href@noop {} {\bibfield  {journal} {\bibinfo  {journal} {Phys.
  Rev. E}\ }\textbf {\bibinfo {volume} {56}},\ \bibinfo {pages} {7166}
  (\bibinfo {year} {1997})}\BibitemShut {NoStop}%
\bibitem [{\citenamefont {Xie}\ and\ \citenamefont {Yu}(2000)}]{xie2000dust}%
  \BibitemOpen
  \bibfield  {author} {\bibinfo {author} {\bibfnamefont {B.~S.}\ \bibnamefont
  {Xie}}\ and\ \bibinfo {author} {\bibfnamefont {M.~Y.}\ \bibnamefont {Yu}},\
  }\href@noop {} {\bibfield  {journal} {\bibinfo  {journal} {Phys. Rev. E}\
  }\textbf {\bibinfo {volume} {62}},\ \bibinfo {pages} {8501} (\bibinfo {year}
  {2000})}\BibitemShut {NoStop}%
\bibitem [{\citenamefont {Anowar}\ \emph {et~al.}(2009)\citenamefont {Anowar},
  \citenamefont {Rahman},\ and\ \citenamefont {Mamun}}]{anowar2009nonlinear}%
  \BibitemOpen
  \bibfield  {author} {\bibinfo {author} {\bibfnamefont {M.~G.~M.}\
  \bibnamefont {Anowar}}, \bibinfo {author} {\bibfnamefont {M.~S.}\
  \bibnamefont {Rahman}}, \ and\ \bibinfo {author} {\bibfnamefont {A.~A.}\
  \bibnamefont {Mamun}},\ }\href@noop {} {\bibfield  {journal} {\bibinfo
  {journal} {Phys. Plasmas}\ }\textbf {\bibinfo {volume} {16}},\ \bibinfo
  {pages} {053704} (\bibinfo {year} {2009})}\BibitemShut {NoStop}%
\bibitem [{\citenamefont {Yaroshenko}\ \emph {et~al.}(2010)\citenamefont
  {Yaroshenko}, \citenamefont {Nosenko}, \citenamefont {Hellberg},
  \citenamefont {Verheest}, \citenamefont {Thomas},\ and\ \citenamefont
  {Morfill}}]{yaroshenko2010nonlinear}%
  \BibitemOpen
  \bibfield  {author} {\bibinfo {author} {\bibfnamefont {V.~V.}\ \bibnamefont
  {Yaroshenko}}, \bibinfo {author} {\bibfnamefont {V.}~\bibnamefont {Nosenko}},
  \bibinfo {author} {\bibfnamefont {M.~A.}\ \bibnamefont {Hellberg}}, \bibinfo
  {author} {\bibfnamefont {F.}~\bibnamefont {Verheest}}, \bibinfo {author}
  {\bibfnamefont {H.~M.}\ \bibnamefont {Thomas}}, \ and\ \bibinfo {author}
  {\bibfnamefont {G.~E.}\ \bibnamefont {Morfill}},\ }\href@noop {} {\bibfield
  {journal} {\bibinfo  {journal} {New J. Phys.}\ }\textbf {\bibinfo {volume}
  {12}},\ \bibinfo {pages} {073038} (\bibinfo {year} {2010})}\BibitemShut
  {NoStop}%
\bibitem [{\citenamefont {Wang}\ \emph {et~al.}(2016)\citenamefont {Wang},
  \citenamefont {Guo},\ and\ \citenamefont {Li}}]{wang2016nonlinear}%
  \BibitemOpen
  \bibfield  {author} {\bibinfo {author} {\bibfnamefont {Y.~L.}\ \bibnamefont
  {Wang}}, \bibinfo {author} {\bibfnamefont {X.~Y.}\ \bibnamefont {Guo}}, \
  and\ \bibinfo {author} {\bibfnamefont {Q.~S.}\ \bibnamefont {Li}},\
  }\href@noop {} {\bibfield  {journal} {\bibinfo  {journal} {Commun. Theor.
  Phys.}\ }\textbf {\bibinfo {volume} {65}},\ \bibinfo {pages} {247} (\bibinfo
  {year} {2016})}\BibitemShut {NoStop}%
\bibitem [{\citenamefont {Quinn}\ and\ \citenamefont
  {Goree}(2000)}]{quinn2000experimental}%
  \BibitemOpen
  \bibfield  {author} {\bibinfo {author} {\bibfnamefont {R.~A.}\ \bibnamefont
  {Quinn}}\ and\ \bibinfo {author} {\bibfnamefont {J.}~\bibnamefont {Goree}},\
  }\href@noop {} {\bibfield  {journal} {\bibinfo  {journal} {Phys. Plasmas}\
  }\textbf {\bibinfo {volume} {7}},\ \bibinfo {pages} {3904} (\bibinfo {year}
  {2000})}\BibitemShut {NoStop}%
\bibitem [{\citenamefont {Amin}\ \emph
  {et~al.}(1998{\natexlab{a}})\citenamefont {Amin}, \citenamefont {Morfill},\
  and\ \citenamefont {Shukla}}]{amin1998modulational}%
  \BibitemOpen
  \bibfield  {author} {\bibinfo {author} {\bibfnamefont {M.~R.}\ \bibnamefont
  {Amin}}, \bibinfo {author} {\bibfnamefont {G.~E.}\ \bibnamefont {Morfill}}, \
  and\ \bibinfo {author} {\bibfnamefont {P.~K.}\ \bibnamefont {Shukla}},\
  }\href@noop {} {\bibfield  {journal} {\bibinfo  {journal} {Phys. Rev. E}\
  }\textbf {\bibinfo {volume} {58}},\ \bibinfo {pages} {6517} (\bibinfo {year}
  {1998}{\natexlab{a}})}\BibitemShut {NoStop}%
\bibitem [{\citenamefont {Kourakis}\ and\ \citenamefont
  {Shukla}(2003)}]{kourakis2003modulational}%
  \BibitemOpen
  \bibfield  {author} {\bibinfo {author} {\bibfnamefont {I.}~\bibnamefont
  {Kourakis}}\ and\ \bibinfo {author} {\bibfnamefont {P.~K.}\ \bibnamefont
  {Shukla}},\ }\href@noop {} {\bibfield  {journal} {\bibinfo  {journal} {Phys.
  Plasmas}\ }\textbf {\bibinfo {volume} {10}},\ \bibinfo {pages} {3459}
  (\bibinfo {year} {2003})}\BibitemShut {NoStop}%
\bibitem [{\citenamefont {Kourakis}\ and\ \citenamefont
  {Shukla}(2004)}]{kourakis2004oblique}%
  \BibitemOpen
  \bibfield  {author} {\bibinfo {author} {\bibfnamefont {I.}~\bibnamefont
  {Kourakis}}\ and\ \bibinfo {author} {\bibfnamefont {P.~K.}\ \bibnamefont
  {Shukla}},\ }\href@noop {} {\bibfield  {journal} {\bibinfo  {journal} {Phys.
  Scr.}\ }\textbf {\bibinfo {volume} {69}},\ \bibinfo {pages} {316} (\bibinfo
  {year} {2004})}\BibitemShut {NoStop}%
\bibitem [{\citenamefont {Duan}\ \emph {et~al.}(2004)\citenamefont {Duan},
  \citenamefont {Parkes},\ and\ \citenamefont {Zhang}}]{duan2004modulational}%
  \BibitemOpen
  \bibfield  {author} {\bibinfo {author} {\bibfnamefont {W.~s.}\ \bibnamefont
  {Duan}}, \bibinfo {author} {\bibfnamefont {J.}~\bibnamefont {Parkes}}, \ and\
  \bibinfo {author} {\bibfnamefont {L.}~\bibnamefont {Zhang}},\ }\href@noop {}
  {\bibfield  {journal} {\bibinfo  {journal} {Phys. Plasmas}\ }\textbf
  {\bibinfo {volume} {11}},\ \bibinfo {pages} {3762} (\bibinfo {year}
  {2004})}\BibitemShut {NoStop}%
\bibitem [{\citenamefont {Misra}\ and\ \citenamefont
  {Chowdhury}(2006)}]{misra2006modulational}%
  \BibitemOpen
  \bibfield  {author} {\bibinfo {author} {\bibfnamefont {A.~P.}\ \bibnamefont
  {Misra}}\ and\ \bibinfo {author} {\bibfnamefont {A.~R.}\ \bibnamefont
  {Chowdhury}},\ }\href@noop {} {\bibfield  {journal} {\bibinfo  {journal}
  {Eur. Phys. J. D}\ }\textbf {\bibinfo {volume} {39}},\ \bibinfo {pages} {49}
  (\bibinfo {year} {2006})}\BibitemShut {NoStop}%
\bibitem [{\citenamefont {El-Taibany}\ and\ \citenamefont
  {Kourakis}(2006)}]{el2006modulational}%
  \BibitemOpen
  \bibfield  {author} {\bibinfo {author} {\bibfnamefont {W.~F.}\ \bibnamefont
  {El-Taibany}}\ and\ \bibinfo {author} {\bibfnamefont {I.}~\bibnamefont
  {Kourakis}},\ }\href@noop {} {\bibfield  {journal} {\bibinfo  {journal}
  {Phys. plasmas}\ }\textbf {\bibinfo {volume} {13}},\ \bibinfo {pages}
  {062302} (\bibinfo {year} {2006})}\BibitemShut {NoStop}%
\bibitem [{\citenamefont {Gill}\ \emph {et~al.}(2010)\citenamefont {Gill},
  \citenamefont {Bains},\ and\ \citenamefont {Bedi}}]{gill2010modulational}%
  \BibitemOpen
  \bibfield  {author} {\bibinfo {author} {\bibfnamefont {T.~S.}\ \bibnamefont
  {Gill}}, \bibinfo {author} {\bibfnamefont {A.~S.}\ \bibnamefont {Bains}}, \
  and\ \bibinfo {author} {\bibfnamefont {C.}~\bibnamefont {Bedi}},\ }\href@noop
  {} {\bibfield  {journal} {\bibinfo  {journal} {Phys. Plasmas}\ }\textbf
  {\bibinfo {volume} {17}},\ \bibinfo {pages} {013701} (\bibinfo {year}
  {2010})}\BibitemShut {NoStop}%
\bibitem [{\citenamefont {Bains}\ \emph {et~al.}(2013)\citenamefont {Bains},
  \citenamefont {Tribeche},\ and\ \citenamefont {Ng}}]{bains2013dust}%
  \BibitemOpen
  \bibfield  {author} {\bibinfo {author} {\bibfnamefont {A.~S.}\ \bibnamefont
  {Bains}}, \bibinfo {author} {\bibfnamefont {M.}~\bibnamefont {Tribeche}}, \
  and\ \bibinfo {author} {\bibfnamefont {C.~S.}\ \bibnamefont {Ng}},\
  }\href@noop {} {\bibfield  {journal} {\bibinfo  {journal} {Astrophys. Space
  Sci.}\ }\textbf {\bibinfo {volume} {343}},\ \bibinfo {pages} {621} (\bibinfo
  {year} {2013})}\BibitemShut {NoStop}%
\bibitem [{\citenamefont {Khaled}\ \emph {et~al.}(2021)\citenamefont {Khaled},
  \citenamefont {Shukri},\ and\ \citenamefont
  {Al-Shaibani}}]{khaled2021modulational}%
  \BibitemOpen
  \bibfield  {author} {\bibinfo {author} {\bibfnamefont {M.~A.~H.}\
  \bibnamefont {Khaled}}, \bibinfo {author} {\bibfnamefont {M.~A.}\
  \bibnamefont {Shukri}}, \ and\ \bibinfo {author} {\bibfnamefont {A.~A.}\
  \bibnamefont {Al-Shaibani}},\ }\href@noop {} {\bibfield  {journal} {\bibinfo
  {journal} {Braz. J. Phys.}\ }\textbf {\bibinfo {volume} {51}},\ \bibinfo
  {pages} {1290} (\bibinfo {year} {2021})}\BibitemShut {NoStop}%
\bibitem [{\citenamefont {Amin}\ \emph
  {et~al.}(1998{\natexlab{b}})\citenamefont {Amin}, \citenamefont {Morfill},\
  and\ \citenamefont {Shukla}}]{amin1998amplitude}%
  \BibitemOpen
  \bibfield  {author} {\bibinfo {author} {\bibfnamefont {M.~R.}\ \bibnamefont
  {Amin}}, \bibinfo {author} {\bibfnamefont {G.~E.}\ \bibnamefont {Morfill}}, \
  and\ \bibinfo {author} {\bibfnamefont {P.~K.}\ \bibnamefont {Shukla}},\
  }\href@noop {} {\bibfield  {journal} {\bibinfo  {journal} {Phys. Plasmas}\
  }\textbf {\bibinfo {volume} {5}},\ \bibinfo {pages} {2578} (\bibinfo {year}
  {1998}{\natexlab{b}})}\BibitemShut {NoStop}%
\bibitem [{\citenamefont {Amin}\ \emph
  {et~al.}(1998{\natexlab{c}})\citenamefont {Amin}, \citenamefont {Morfill},\
  and\ \citenamefont {Shukla}}]{amin1998modulational_dustlattice}%
  \BibitemOpen
  \bibfield  {author} {\bibinfo {author} {\bibfnamefont {M.~R.}\ \bibnamefont
  {Amin}}, \bibinfo {author} {\bibfnamefont {G.~E.}\ \bibnamefont {Morfill}}, \
  and\ \bibinfo {author} {\bibfnamefont {P.~K.}\ \bibnamefont {Shukla}},\
  }\href@noop {} {\bibfield  {journal} {\bibinfo  {journal} {Phys. Scr.}\
  }\textbf {\bibinfo {volume} {58}},\ \bibinfo {pages} {628} (\bibinfo {year}
  {1998}{\natexlab{c}})}\BibitemShut {NoStop}%
\bibitem [{\citenamefont {Kourakis}\ and\ \citenamefont
  {Shukla}(2006)}]{kourakis2006nonlinear}%
  \BibitemOpen
  \bibfield  {author} {\bibinfo {author} {\bibfnamefont {I.}~\bibnamefont
  {Kourakis}}\ and\ \bibinfo {author} {\bibfnamefont {P.~K.}\ \bibnamefont
  {Shukla}},\ }\href@noop {} {\bibfield  {journal} {\bibinfo  {journal} {Int.
  J. Bifurcation Chaos}\ }\textbf {\bibinfo {volume} {16}},\ \bibinfo {pages}
  {1711} (\bibinfo {year} {2006})}\BibitemShut {NoStop}%
\bibitem [{\citenamefont {Sultana}(2020)}]{sultana2020dust}%
  \BibitemOpen
  \bibfield  {author} {\bibinfo {author} {\bibfnamefont {S.}~\bibnamefont
  {Sultana}},\ }\href@noop {} {\bibfield  {journal} {\bibinfo  {journal} {Eur.
  Phys. J. D}\ }\textbf {\bibinfo {volume} {74}},\ \bibinfo {pages} {1}
  (\bibinfo {year} {2020})}\BibitemShut {NoStop}%
\bibitem [{\citenamefont {Ikezi}(1986)}]{ikezi1986coulomb}%
  \BibitemOpen
  \bibfield  {author} {\bibinfo {author} {\bibfnamefont {H.}~\bibnamefont
  {Ikezi}},\ }\href@noop {} {\bibfield  {journal} {\bibinfo  {journal} {Phys.
  Fluids}\ }\textbf {\bibinfo {volume} {29}},\ \bibinfo {pages} {1764}
  (\bibinfo {year} {1986})}\BibitemShut {NoStop}%
\bibitem [{\citenamefont {Thomas}\ \emph {et~al.}(1994)\citenamefont {Thomas},
  \citenamefont {Morfill}, \citenamefont {Demmel}, \citenamefont {Goree},
  \citenamefont {Feuerbacher},\ and\ \citenamefont
  {M{\"o}hlmann}}]{thomas1994plasma}%
  \BibitemOpen
  \bibfield  {author} {\bibinfo {author} {\bibfnamefont {H.}~\bibnamefont
  {Thomas}}, \bibinfo {author} {\bibfnamefont {G.~E.}\ \bibnamefont {Morfill}},
  \bibinfo {author} {\bibfnamefont {V.}~\bibnamefont {Demmel}}, \bibinfo
  {author} {\bibfnamefont {J.}~\bibnamefont {Goree}}, \bibinfo {author}
  {\bibfnamefont {B.}~\bibnamefont {Feuerbacher}}, \ and\ \bibinfo {author}
  {\bibfnamefont {D.}~\bibnamefont {M{\"o}hlmann}},\ }\href@noop {} {\bibfield
  {journal} {\bibinfo  {journal} {Phys. Rev. Lett.}\ }\textbf {\bibinfo
  {volume} {73}},\ \bibinfo {pages} {652} (\bibinfo {year} {1994})}\BibitemShut
  {NoStop}%
\bibitem [{\citenamefont {Chu}\ and\ \citenamefont
  {Lin}(1994)}]{chu1994direct}%
  \BibitemOpen
  \bibfield  {author} {\bibinfo {author} {\bibfnamefont {J.~H.}\ \bibnamefont
  {Chu}}\ and\ \bibinfo {author} {\bibfnamefont {I.}~\bibnamefont {Lin}},\
  }\href@noop {} {\bibfield  {journal} {\bibinfo  {journal} {Phys. Rev lett.}\
  }\textbf {\bibinfo {volume} {72}},\ \bibinfo {pages} {4009} (\bibinfo {year}
  {1994})}\BibitemShut {NoStop}%
\bibitem [{\citenamefont {Misawa}\ \emph {et~al.}(2001)\citenamefont {Misawa},
  \citenamefont {Ohno}, \citenamefont {Asano}, \citenamefont {Sawai},
  \citenamefont {Takamura},\ and\ \citenamefont
  {Kaw}}]{misawa2001experimental}%
  \BibitemOpen
  \bibfield  {author} {\bibinfo {author} {\bibfnamefont {T.}~\bibnamefont
  {Misawa}}, \bibinfo {author} {\bibfnamefont {N.}~\bibnamefont {Ohno}},
  \bibinfo {author} {\bibfnamefont {K.}~\bibnamefont {Asano}}, \bibinfo
  {author} {\bibfnamefont {M.}~\bibnamefont {Sawai}}, \bibinfo {author}
  {\bibfnamefont {S.}~\bibnamefont {Takamura}}, \ and\ \bibinfo {author}
  {\bibfnamefont {P.~K.}\ \bibnamefont {Kaw}},\ }\href@noop {} {\bibfield
  {journal} {\bibinfo  {journal} {Phys. Rev. Lett.}\ }\textbf {\bibinfo
  {volume} {86}},\ \bibinfo {pages} {1219} (\bibinfo {year}
  {2001})}\BibitemShut {NoStop}%
\bibitem [{\citenamefont {Xie}\ \emph {et~al.}(2002)\citenamefont {Xie},
  \citenamefont {Yu}, \citenamefont {He}, \citenamefont {Chen},\ and\
  \citenamefont {Liu}}]{xie2002modulational}%
  \BibitemOpen
  \bibfield  {author} {\bibinfo {author} {\bibfnamefont {B.~S.}\ \bibnamefont
  {Xie}}, \bibinfo {author} {\bibfnamefont {M.~Y.}\ \bibnamefont {Yu}},
  \bibinfo {author} {\bibfnamefont {K.~F.}\ \bibnamefont {He}}, \bibinfo
  {author} {\bibfnamefont {Z.~Y.}\ \bibnamefont {Chen}}, \ and\ \bibinfo
  {author} {\bibfnamefont {S.~B.}\ \bibnamefont {Liu}},\ }\href@noop {}
  {\bibfield  {journal} {\bibinfo  {journal} {Phys. Rev. E}\ }\textbf {\bibinfo
  {volume} {65}},\ \bibinfo {pages} {027401} (\bibinfo {year}
  {2002})}\BibitemShut {NoStop}%
\bibitem [{\citenamefont {Chaudhuri}\ \emph {et~al.}(2019)\citenamefont
  {Chaudhuri}, \citenamefont {Chowdhury},\ and\ \citenamefont
  {Chowdhury}}]{chaudhuri2019solitary}%
  \BibitemOpen
  \bibfield  {author} {\bibinfo {author} {\bibfnamefont {S.}~\bibnamefont
  {Chaudhuri}}, \bibinfo {author} {\bibfnamefont {K.~R.}\ \bibnamefont
  {Chowdhury}}, \ and\ \bibinfo {author} {\bibfnamefont {A.~R.}\ \bibnamefont
  {Chowdhury}},\ }\href@noop {} {\bibfield  {journal} {\bibinfo  {journal}
  {Pramana}\ }\textbf {\bibinfo {volume} {92}},\ \bibinfo {pages} {1} (\bibinfo
  {year} {2019})}\BibitemShut {NoStop}%
\bibitem [{\citenamefont {El-Labany}\ \emph {et~al.}(2015)\citenamefont
  {El-Labany}, \citenamefont {El-Shamy}, \citenamefont {El~Taibany},\ and\
  \citenamefont {Zedan}}]{el2015modeling}%
  \BibitemOpen
  \bibfield  {author} {\bibinfo {author} {\bibfnamefont {S.~K.}\ \bibnamefont
  {El-Labany}}, \bibinfo {author} {\bibfnamefont {E.~F.}\ \bibnamefont
  {El-Shamy}}, \bibinfo {author} {\bibfnamefont {W.~F.}\ \bibnamefont
  {El~Taibany}}, \ and\ \bibinfo {author} {\bibfnamefont {N.~A.}\ \bibnamefont
  {Zedan}},\ }\href@noop {} {\bibfield  {journal} {\bibinfo  {journal} {Chin.
  Phys. B}\ }\textbf {\bibinfo {volume} {24}},\ \bibinfo {pages} {035201}
  (\bibinfo {year} {2015})}\BibitemShut {NoStop}%
\bibitem [{\citenamefont {Kalman}\ and\ \citenamefont
  {Rosenberg}(2003)}]{kalman2003instabilities}%
  \BibitemOpen
  \bibfield  {author} {\bibinfo {author} {\bibfnamefont {G.~J.}\ \bibnamefont
  {Kalman}}\ and\ \bibinfo {author} {\bibfnamefont {M.}~\bibnamefont
  {Rosenberg}},\ }\href@noop {} {\bibfield  {journal} {\bibinfo  {journal} {J.
  Phys. A: Math. Gen.}\ }\textbf {\bibinfo {volume} {36}},\ \bibinfo {pages}
  {5963} (\bibinfo {year} {2003})}\BibitemShut {NoStop}%
\bibitem [{\citenamefont {Rosenberg}\ \emph {et~al.}(2012)\citenamefont
  {Rosenberg}, \citenamefont {Kalman},\ and\ \citenamefont
  {Hartmann}}]{rosenberg2012instabilities}%
  \BibitemOpen
  \bibfield  {author} {\bibinfo {author} {\bibfnamefont {M.}~\bibnamefont
  {Rosenberg}}, \bibinfo {author} {\bibfnamefont {G.~J.}\ \bibnamefont
  {Kalman}}, \ and\ \bibinfo {author} {\bibfnamefont {P.}~\bibnamefont
  {Hartmann}},\ }\href@noop {} {\bibfield  {journal} {\bibinfo  {journal}
  {Contr. Plasma Phys.}\ }\textbf {\bibinfo {volume} {52}},\ \bibinfo {pages}
  {70} (\bibinfo {year} {2012})}\BibitemShut {NoStop}%
\bibitem [{\citenamefont {Rosenberg}\ \emph {et~al.}(2014)\citenamefont
  {Rosenberg}, \citenamefont {Kalman}, \citenamefont {Hartmann},\ and\
  \citenamefont {Goree}}]{rosenberg2014effect}%
  \BibitemOpen
  \bibfield  {author} {\bibinfo {author} {\bibfnamefont {M.}~\bibnamefont
  {Rosenberg}}, \bibinfo {author} {\bibfnamefont {G.~J.}\ \bibnamefont
  {Kalman}}, \bibinfo {author} {\bibfnamefont {P.}~\bibnamefont {Hartmann}}, \
  and\ \bibinfo {author} {\bibfnamefont {J.}~\bibnamefont {Goree}},\
  }\href@noop {} {\bibfield  {journal} {\bibinfo  {journal} {Phys. Rev. E}\
  }\textbf {\bibinfo {volume} {89}},\ \bibinfo {pages} {013103} (\bibinfo
  {year} {2014})}\BibitemShut {NoStop}%
\bibitem [{\citenamefont {Zhou}\ and\ \citenamefont
  {Ouyang}(2021)}]{zhou2021mass}%
  \BibitemOpen
  \bibfield  {author} {\bibinfo {author} {\bibfnamefont {X.}~\bibnamefont
  {Zhou}}\ and\ \bibinfo {author} {\bibfnamefont {Z.}~\bibnamefont {Ouyang}},\
  }\href@noop {} {\bibfield  {journal} {\bibinfo  {journal} {Anal. Chem.}\
  }\textbf {\bibinfo {volume} {93}},\ \bibinfo {pages} {5998} (\bibinfo {year}
  {2021})}\BibitemShut {NoStop}%
\bibitem [{\citenamefont {Chamel}\ \emph {et~al.}(2016)\citenamefont {Chamel},
  \citenamefont {Page},\ and\ \citenamefont {Reddy}}]{chamel2016collective}%
  \BibitemOpen
  \bibfield  {author} {\bibinfo {author} {\bibfnamefont {N.}~\bibnamefont
  {Chamel}}, \bibinfo {author} {\bibfnamefont {D.}~\bibnamefont {Page}}, \ and\
  \bibinfo {author} {\bibfnamefont {S.}~\bibnamefont {Reddy}},\ }in\ \href@noop
  {} {\emph {\bibinfo {booktitle} {J. Phys.: Conf. Ser.}}},\ Vol.\ \bibinfo
  {volume} {665}\ (\bibinfo {organization} {IOP Publishing},\ \bibinfo {year}
  {2016})\ p.\ \bibinfo {pages} {012065}\BibitemShut {NoStop}%
\bibitem [{\citenamefont {Stacey}(2007)}]{stacey2007survey}%
  \BibitemOpen
  \bibfield  {author} {\bibinfo {author} {\bibfnamefont {W.~M.}\ \bibnamefont
  {Stacey}},\ }\href@noop {} {\bibfield  {journal} {\bibinfo  {journal} {Fusion
  science and technology}\ }\textbf {\bibinfo {volume} {52}},\ \bibinfo {pages}
  {29} (\bibinfo {year} {2007})}\BibitemShut {NoStop}%
\bibitem [{\citenamefont {Killian}\ \emph {et~al.}(2007)\citenamefont
  {Killian}, \citenamefont {Pattard}, \citenamefont {Pohl},\ and\ \citenamefont
  {Rost}}]{killian2007ultracold}%
  \BibitemOpen
  \bibfield  {author} {\bibinfo {author} {\bibfnamefont {T.~C.}\ \bibnamefont
  {Killian}}, \bibinfo {author} {\bibfnamefont {T.}~\bibnamefont {Pattard}},
  \bibinfo {author} {\bibfnamefont {T.}~\bibnamefont {Pohl}}, \ and\ \bibinfo
  {author} {\bibfnamefont {J.}~\bibnamefont {Rost}},\ }\href@noop {} {\bibfield
   {journal} {\bibinfo  {journal} {Phys. Rep.}\ }\textbf {\bibinfo {volume}
  {449}},\ \bibinfo {pages} {77} (\bibinfo {year} {2007})}\BibitemShut
  {NoStop}%
\bibitem [{\citenamefont {Lyon}\ and\ \citenamefont
  {Rolston}(2016)}]{lyon2016ultracold}%
  \BibitemOpen
  \bibfield  {author} {\bibinfo {author} {\bibfnamefont {M.}~\bibnamefont
  {Lyon}}\ and\ \bibinfo {author} {\bibfnamefont {S.}~\bibnamefont {Rolston}},\
  }\href@noop {} {\bibfield  {journal} {\bibinfo  {journal} {Rep. Prog. Phys.}\
  }\textbf {\bibinfo {volume} {80}},\ \bibinfo {pages} {017001} (\bibinfo
  {year} {2016})}\BibitemShut {NoStop}%
\bibitem [{\citenamefont {Taniuti}\ and\ \citenamefont
  {Yajima}(1969)}]{taniuti1969perturbation}%
  \BibitemOpen
  \bibfield  {author} {\bibinfo {author} {\bibfnamefont {T.}~\bibnamefont
  {Taniuti}}\ and\ \bibinfo {author} {\bibfnamefont {N.}~\bibnamefont
  {Yajima}},\ }\href@noop {} {\bibfield  {journal} {\bibinfo  {journal} {J.
  Math. Phys.}\ }\textbf {\bibinfo {volume} {10}},\ \bibinfo {pages} {1369}
  (\bibinfo {year} {1969})}\BibitemShut {NoStop}%
\bibitem [{\citenamefont {Asano}\ \emph {et~al.}(1969)\citenamefont {Asano},
  \citenamefont {Taniuti},\ and\ \citenamefont
  {Yajima}}]{asano1969perturbation}%
  \BibitemOpen
  \bibfield  {author} {\bibinfo {author} {\bibfnamefont {N.}~\bibnamefont
  {Asano}}, \bibinfo {author} {\bibfnamefont {T.}~\bibnamefont {Taniuti}}, \
  and\ \bibinfo {author} {\bibfnamefont {N.}~\bibnamefont {Yajima}},\
  }\href@noop {} {\bibfield  {journal} {\bibinfo  {journal} {J. Math. Phys.}\
  }\textbf {\bibinfo {volume} {10}},\ \bibinfo {pages} {2020} (\bibinfo {year}
  {1969})}\BibitemShut {NoStop}%
\bibitem [{\citenamefont {Kumar}\ and\ \citenamefont
  {Sharma}(2021)}]{kumar2021collective}%
  \BibitemOpen
  \bibfield  {author} {\bibinfo {author} {\bibfnamefont {P.}~\bibnamefont
  {Kumar}}\ and\ \bibinfo {author} {\bibfnamefont {D.}~\bibnamefont {Sharma}},\
  }\href@noop {} {\bibfield  {journal} {\bibinfo  {journal} {Phys. Plasmas}\
  }\textbf {\bibinfo {volume} {28}},\ \bibinfo {pages} {083704} (\bibinfo
  {year} {2021})}\BibitemShut {NoStop}%
\bibitem [{\citenamefont {Golden}\ and\ \citenamefont
  {Kalman}(2000)}]{golden2000quasilocalized}%
  \BibitemOpen
  \bibfield  {author} {\bibinfo {author} {\bibfnamefont {K.~I.}\ \bibnamefont
  {Golden}}\ and\ \bibinfo {author} {\bibfnamefont {G.~J.}\ \bibnamefont
  {Kalman}},\ }\href@noop {} {\bibfield  {journal} {\bibinfo  {journal}
  {Phys.Plasmas}\ }\textbf {\bibinfo {volume} {7}},\ \bibinfo {pages} {14}
  (\bibinfo {year} {2000})}\BibitemShut {NoStop}%
\bibitem [{\citenamefont {Hou}\ \emph {et~al.}(2004)\citenamefont {Hou},
  \citenamefont {Wang},\ and\ \citenamefont
  {Mi{\v{s}}kovi{\'c}}}]{hou2004theoretical}%
  \BibitemOpen
  \bibfield  {author} {\bibinfo {author} {\bibfnamefont {L.~J.}\ \bibnamefont
  {Hou}}, \bibinfo {author} {\bibfnamefont {Y.~N.}\ \bibnamefont {Wang}}, \
  and\ \bibinfo {author} {\bibfnamefont {Z.~L.}\ \bibnamefont
  {Mi{\v{s}}kovi{\'c}}},\ }\href@noop {} {\bibfield  {journal} {\bibinfo
  {journal} {Phys. Rev. E}\ }\textbf {\bibinfo {volume} {70}},\ \bibinfo
  {pages} {056406} (\bibinfo {year} {2004})}\BibitemShut {NoStop}%
\bibitem [{\citenamefont {Hou}\ \emph {et~al.}(2009)\citenamefont {Hou},
  \citenamefont {Mi{\v{s}}kovi{\'c}}, \citenamefont {Piel},\ and\ \citenamefont
  {Murillo}}]{hou2009wave}%
  \BibitemOpen
  \bibfield  {author} {\bibinfo {author} {\bibfnamefont {L.~J.}\ \bibnamefont
  {Hou}}, \bibinfo {author} {\bibfnamefont {Z.~L.}\ \bibnamefont
  {Mi{\v{s}}kovi{\'c}}}, \bibinfo {author} {\bibfnamefont {A.}~\bibnamefont
  {Piel}}, \ and\ \bibinfo {author} {\bibfnamefont {M.~S.}\ \bibnamefont
  {Murillo}},\ }\href@noop {} {\bibfield  {journal} {\bibinfo  {journal} {Phys.
  Rev. E}\ }\textbf {\bibinfo {volume} {79}},\ \bibinfo {pages} {046412}
  (\bibinfo {year} {2009})}\BibitemShut {NoStop}%
\bibitem [{\citenamefont {Lado}(1978)}]{lado1978hypernetted}%
  \BibitemOpen
  \bibfield  {author} {\bibinfo {author} {\bibfnamefont {F.}~\bibnamefont
  {Lado}},\ }\href@noop {} {\bibfield  {journal} {\bibinfo  {journal} {Phys.
  Rev. B}\ }\textbf {\bibinfo {volume} {17}},\ \bibinfo {pages} {2827}
  (\bibinfo {year} {1978})}\BibitemShut {NoStop}%
\bibitem [{\citenamefont {Hartmann}\ \emph {et~al.}(2005)\citenamefont
  {Hartmann}, \citenamefont {Kalman}, \citenamefont {Donk{\'o}},\ and\
  \citenamefont {Kutasi}}]{hartmann2005equilibrium}%
  \BibitemOpen
  \bibfield  {author} {\bibinfo {author} {\bibfnamefont {P.}~\bibnamefont
  {Hartmann}}, \bibinfo {author} {\bibfnamefont {G.}~\bibnamefont {Kalman}},
  \bibinfo {author} {\bibfnamefont {Z.}~\bibnamefont {Donk{\'o}}}, \ and\
  \bibinfo {author} {\bibfnamefont {K.}~\bibnamefont {Kutasi}},\ }\href@noop {}
  {\bibfield  {journal} {\bibinfo  {journal} {Phys. Rev. E}\ }\textbf {\bibinfo
  {volume} {72}},\ \bibinfo {pages} {026409} (\bibinfo {year}
  {2005})}\BibitemShut {NoStop}%
\bibitem [{\citenamefont {Khrapak}\ \emph {et~al.}(2016)\citenamefont
  {Khrapak}, \citenamefont {Klumov}, \citenamefont {Couedel},\ and\
  \citenamefont {Thomas}}]{khrapak2016long}%
  \BibitemOpen
  \bibfield  {author} {\bibinfo {author} {\bibfnamefont {S.~A.}\ \bibnamefont
  {Khrapak}}, \bibinfo {author} {\bibfnamefont {B.}~\bibnamefont {Klumov}},
  \bibinfo {author} {\bibfnamefont {L.}~\bibnamefont {Couedel}}, \ and\
  \bibinfo {author} {\bibfnamefont {H.~M.}\ \bibnamefont {Thomas}},\
  }\href@noop {} {\bibfield  {journal} {\bibinfo  {journal} {Phys. Plasmas}\
  }\textbf {\bibinfo {volume} {23}},\ \bibinfo {pages} {023702} (\bibinfo
  {year} {2016})}\BibitemShut {NoStop}%
\bibitem [{\citenamefont {Dalui}\ \emph {et~al.}(2017)\citenamefont {Dalui},
  \citenamefont {Bandyopadhyay},\ and\ \citenamefont
  {Das}}]{dalui2017modulational}%
  \BibitemOpen
  \bibfield  {author} {\bibinfo {author} {\bibfnamefont {S.}~\bibnamefont
  {Dalui}}, \bibinfo {author} {\bibfnamefont {A.}~\bibnamefont
  {Bandyopadhyay}}, \ and\ \bibinfo {author} {\bibfnamefont {K.~P.}\
  \bibnamefont {Das}},\ }\href@noop {} {\bibfield  {journal} {\bibinfo
  {journal} {Phys. Plasmas}\ }\textbf {\bibinfo {volume} {24}},\ \bibinfo
  {pages} {042305} (\bibinfo {year} {2017})}\BibitemShut {NoStop}%
\bibitem [{\citenamefont {Khrapak}(2017)}]{khrapak2017practical}%
  \BibitemOpen
  \bibfield  {author} {\bibinfo {author} {\bibfnamefont {S.~A.}\ \bibnamefont
  {Khrapak}},\ }\href@noop {} {\bibfield  {journal} {\bibinfo  {journal} {AIP
  Advances}\ }\textbf {\bibinfo {volume} {7}},\ \bibinfo {pages} {125026}
  (\bibinfo {year} {2017})}\BibitemShut {NoStop}%
\bibitem [{\citenamefont {Fedele}\ and\ \citenamefont
  {Schamel}(2002)}]{fedele2002solitary}%
  \BibitemOpen
  \bibfield  {author} {\bibinfo {author} {\bibfnamefont {R.}~\bibnamefont
  {Fedele}}\ and\ \bibinfo {author} {\bibfnamefont {H.}~\bibnamefont
  {Schamel}},\ }\href@noop {} {\bibfield  {journal} {\bibinfo  {journal} {Eur.
  Phys. J. B}\ }\textbf {\bibinfo {volume} {27}},\ \bibinfo {pages} {313}
  (\bibinfo {year} {2002})}\BibitemShut {NoStop}%
\bibitem [{\citenamefont {Fedele}(2002)}]{fedele2002envelope}%
  \BibitemOpen
  \bibfield  {author} {\bibinfo {author} {\bibfnamefont {R.}~\bibnamefont
  {Fedele}},\ }\href@noop {} {\bibfield  {journal} {\bibinfo  {journal} {Phys.
  Scr.}\ }\textbf {\bibinfo {volume} {65}},\ \bibinfo {pages} {502} (\bibinfo
  {year} {2002})}\BibitemShut {NoStop}%
\bibitem [{\citenamefont {Sikdar}\ \emph {et~al.}(2018)\citenamefont {Sikdar},
  \citenamefont {Adak}, \citenamefont {Ghosh},\ and\ \citenamefont
  {Khan}}]{sikdar2018electrostatic}%
  \BibitemOpen
  \bibfield  {author} {\bibinfo {author} {\bibfnamefont {A.}~\bibnamefont
  {Sikdar}}, \bibinfo {author} {\bibfnamefont {A.}~\bibnamefont {Adak}},
  \bibinfo {author} {\bibfnamefont {S.}~\bibnamefont {Ghosh}}, \ and\ \bibinfo
  {author} {\bibfnamefont {M.}~\bibnamefont {Khan}},\ }\href@noop {} {\bibfield
   {journal} {\bibinfo  {journal} {Phys. Plasmas}\ }\textbf {\bibinfo {volume}
  {25}},\ \bibinfo {pages} {052303} (\bibinfo {year} {2018})}\BibitemShut
  {NoStop}%
\end{thebibliography}%
%\bibliography{paper_2.bib}

%\end{thebibliography}

%%%%%%%%%%%%%%%%%%%%%%%%%%%%%%%%%%%%%%%%%%%%%%%%%%%%%%%%%%%%%%%%%%%%%%%
\end{document}